\documentclass[preprint,12pt]{elsarticle}

\usepackage{graphicx}
\usepackage{fancyhdr}
\usepackage{lineno}
\usepackage{amssymb}
\usepackage{amsthm}
\usepackage{caption}
\usepackage{subcaption}
\usepackage{booktabs}
\usepackage{tabularx}
\usepackage{amsmath}
\usepackage{dsfont}
\usepackage{multicol}
\usepackage{mathtools}
\usepackage{float}
\usepackage{multirow}
\usepackage{natbib}
\usepackage{color}
\restylefloat{table}
\usepackage{upgreek}
\usepackage{nicefrac}
\usepackage{xargs}                      
\usepackage[pdftex,dvipsnames]{xcolor} 
\usepackage[colorinlistoftodos,prependcaption,textsize=tiny]{todonotes}
\usepackage{xcolor}
\usepackage{multicol}
\usepackage{algorithm}
\usepackage{bm}
\usepackage[noend]{algpseudocode}
\makeatletter
\def\BState{\State\hskip-\ALG@thistlm}
\makeatother
\usepackage{nomencl}
\usepackage{url}
\usepackage{subcaption}
\usepackage[margin=2.5cm]{geometry}

\renewcommand{\vec}[1]{\mathbf{#1}}

\journal{Journal Name}

\begin{document}

\begin{frontmatter}


\title{Imprecise Subset Simulation}



\author{Dimitris G. Giovanis}
\author{Michael D. Shields}

\address{Department of Civil \& Systems Engineering, Johns Hopkins University, Baltimore, MD, 21218, USA}

\begin{abstract}
The objective of this work is to quantify the uncertainty in probability of failure estimates resulting from incomplete knowledge of the probability distributions for the input random variables. We propose a framework that couples the widely used Subset simulation (SuS) with Bayesian/information theoretic multi-model inference. The process starts with data used to infer probability distributions for the model inputs. Often such data sets are small. Multi-model inference is used to assess uncertainty associated with the model-form and parameters of these random variables in the form of model probabilities and the associated joint parameter probability densities. A sampling procedure is used to construct a set of equally probable candidate probability distributions and an optimal importance sampling distribution is determined analytically from this set. Subset simulation is then performed using this optimal sampling density and the resulting conditional probabilities are re-weighted using importance sampling. The result of this process are empirical probability distributions of failure probabilities that provide direct estimates of the uncertainty in failure probability estimates that result from inference on small data sets. The method is demonstrated to be both computationally efficient -- requiring only a single subset simulation and nominal cost of sample re-weighting -- and to provide reasonable estimates of the uncertainty in failure probabilities.
\end{abstract}


\end{frontmatter}


\section{Introduction}
\noindent
Due to the inherent uncertainty in design parameters, material properties and loading conditions involved in structural analysis, safety measures are necessary to guarantee that a structure will satisfy its intended performance measures. Reliability analysis is the study of structural performance, where performance is classified as either satisfactory (safe) or unsatisfactory (failure).  Assuming that the behavior of a system is described by a performance function  $g(\vec{X})$, where $\vec{X}=[X_1, \ldots, X_n]\in \mathbb{R}^{n}$ is a vector of the $n$ random variables having joint probability density $f(\vec{x})$ characterizing uncertainty in the system, reliability analysis is specifically interested in calculating the probability of failure
\begin{equation}\label{p_f}
P_F =\mathbb{P}(g(\textbf{X}) \leq 0) = \int_{F}f(\vec{x})d\vec{x} = \int \text{I}_{F}(\vec{x})f(\vec{x})\text{d}\vec{x}
\end{equation}
where $F$ is the failure region (region of unsatisfactory performance), and $\text{I}_{F}$ is the indicator function with $\text{I}_{F}(\vec{x})=1$ if $g(\vec{x})\in F$ and $\text{I}_{F}(\vec{x})=0$ otherwise. 
For realistic large-scale systems, estimation of the failure probability with Eq.\eqref{p_f} is challenging for several reasons. Most notably: the performance function ($g(\vec{x})$) is typically computed using a complex numerical model and cannot be expressed in closed-form; the vector of input random variables $\vec{X}$ may be very high dimensional, and the joint density function of random vector $\vec{X}$ may not be precisely known -- especially when distributions are estimated from limited data sets.  This paper specifically deals with conducting reliability analysis in this final case, when the distribution of random variables in the system is uncertain, specifically because they are being inferred from small data sets.

Despite these challenges, a vast number of methods have been developed to accurately and efficiently estimate the reliability of a system. These efforts date back to the 1970s with the pioneering First and Second Order Reliability methods (FORM/SORM) \cite{Rackwitz1978, Breitung1989} that use a Taylor series expansion to approximate the limit-state function $g(\vec{X})=0$ and also include the widely used response surface methods \cite{Myers, Faravelli1989} based on polynomial approximations of the performance function and, of course, classical Monte Carlo simulation-based approaches \cite{Fishman1996, Rubinstein1981}. Among these methods, Monte Carlo methods are the most robust when solving high-dimensional problems and those with complex failure regions. However, Monte Carlo simulation-based methods require a very-large number of performance function evaluations (e.g. finite element analyses) which makes them computationally expensive. This has inspired the development of more sophisticated Monte Carlo simulation-based approaches over the past $\sim20$ years that leverage variance reduction methods such as Latin hypercube sampling \cite{olsson2003latin},  line sampling \cite{Pradlwarter2007}, importance sampling \cite{Pradlwarter2004,Koutsourelakis2004}, and subset simulation \cite{Au2001} that reduce the number of required performance function evaluations while maintaining sufficient accuracy in failure probability estimates. In recent years, a great deal of research has also focused on surrogate model-based approaches (e.g.\ \cite{bichon2008efficient, Echard, Papadopoulos2012,Sundar2016, marelli2018active}) that replace the expensive performance function with an inexpensive mathematical approximation. With these advanced tools, the estimation of the failure probabilities even for many complex nonlinear systems with non-Gaussian uncertainties has become a feasible task.


Despite these advances toward efficiently computing very precise failure probability estimates, we must consider how much confidence we can place in these estimates. That is, although these estimates are highly \textit{precise}, there are additional factors that contribute significant \textit{uncertainty} to these estimators. Chief among those factors is the precision with which distributions for the input random variables are quantified. Distributions for these random variables are typically inferred from real data, with the challenge that the available data are often scarce. Consequently, confidence in the inferred probability distributions may be low, which will translate to low confidence in reliability estimates. How then can we assess the confidence is reliability estimates in the presence of probability distribution uncertainty?

Over the past several years the scientific community has begun to focus on the impact of imperfect knowledge about the probability distributions on structural reliability estimates. Starting in the early 2000s, imprecise reliability analysis has been studied using the various mathematical theories for imprecision in combination with well-known reliability methods \cite{alvarez2006calculation,alvarez2014efficient,ALVAREZ2017101,alvarez2018estimation}. For example, Du \cite{du2008unified}, applied interval theory to assess plausibility and belief functions using FORM. Hurtado et al.\ \cite{hurtado2012fuzzy} used fuzzy sets to estimate the membership function of the response. Others, including de Angelis et al.\ \cite{de2015advanced}, Fetz and Oberguggenberger \cite{fetz2016imprecise}, and Zhang and Shields \cite{Zhang2018} consider ensembles of models and sample reweighting schemes. In more recent works, Wang et al.\ \cite{WANG201892} estimated the bounds of the failure probability directly without constructing the probability bounds of the input random variables, Zhang et al.\ \cite{ZHANG2013137} proposed an interval quasi-Monte Carlo simulation methodology and Zhao et al.\ \cite{ZHAO2018160} proposed a method for estimating the mean and quantile of the probability of failure given uncertainty in the distribution parameters. 

In this work, we address the challenge of quantifying the uncertainty in probability of failure estimates using Subset simulation (SuS) -- a robust and widely-used simulation-based method for reliability analysis.  Our motivation stems from the fact that SuS is very efficient and precise when probability distributions (and hence the conditional levels) are precisely defined. However, when probability models are uncertain, the associated conditional levels for subset simulation also become uncertainty, which complicates sample selection and conditional probability estimates. We propose an approach based on Bayesian/Information Theoretic multi-model selection where first, distribution uncertainty is quantified by identifying a multi-model set of candidate distributions -- each of which is equally probable of represented the limited data used for inference. Using this multi-model set, we analytically identify an optimal sampling density using a method developed in \cite{Zhang2018}. Subset simulation is performed once using this optimal sampling density. We then apply an importance sampling re-weighting scheme to recompute the conditional probabilities for each distribution model in the candidate set. The result of the proposed imprecise Subset Simulation method is an empirical probability distribution of failure probabilities that allows us to estimate the confidence in Subset Simulation results. It is shown that this uncertainty may be large in relation to the probability of failure estimate itself (often several orders of magnitude when data sets are small), meaning that caution should be exercised in placing too much confidence in probability of failure estimates. 

The proposed approach is computationally efficient because it requires only a single Subset Simulation and the nominal computational cost associated with sampling re-weighting. We demonstrate the approach using two engineering examples. The first is a simplified plate buckling problem having uncertainties in material properties derived from historical data. The second consider the uncertainty in a dynamic model of a simple frame structure subjected to base excitation with uncertain stiffness and damping that are inferred from limited data. 


The paper is organized as follows. In Sections 2 and 3, we review the essentials of Subset Simulation and multi-model inference, respectively. The proposed method is introduced in Section 4 and applied to engineering systems in Section 5. Finally, some concluding remarks are provieed in Section 6.

We also note that SuS is employed herein using the affine invariant ensemble sampler as described in \cite{shields2021subset}. All of the tools necessary for implementation of the proposed method, including multi-model inference and SuS are available in the open-source UQpy Python toolbox \cite{olivierUQpy}.

\section{Subset Simulation}

\noindent
Subset simulation (SuS) is a variance reduction method that has drawn significant attention because of its efficiency in estimating small failure probabilities in high-dimensional parameter spaces \cite{Au2001}. In SuS, the failure domain $F = \{\vec{X} \in \mathbb{R}^{n}:g(\vec{X}) \leq 0\}$ is represented as by the intersection of a set of $m$ nested intermediate failure events $F_1 \supset F_2 \supset \cdots \supset F_m=F$. This division into subsets transforms the simulation of rare events into a sequence of simulations of more frequent conditional events and correspondingly expresses the small probability of failure as a product of larger conditional probabilities as
 \begin{equation}\label{eq:Failure_prob}
P_F = \mathbb{P}[F]=\mathbb{P}[\cap_{i=1}^{m} F_i]=\mathbb{P}[F_1]\prod_{i=1}^{m-1} \mathbb{P}(F_{i+1} | F_{i}) = P_1\prod_{i=1}^{m-1} P_i
 \end{equation}
Each intermediate failure event is defined as $F_i=\{G(\vec{X})\le b_i\}$, where $b_1>b_2>\ldots>b_m=0$ are threshold values, usually selected such that each subset has equal conditional probabilities $P_i$ $i=1, \ldots, m$. Additional guidance on the selection of conditional levels and conditional probabilities can be found in \cite{zuev2012bayesian}. 

In this setting, probability $P_1$ can be easily computed through Monte Carlo simulation. 
However, calculating the conditional probabilities $P_i$, $i=2, \ldots, m$ requires the simulation of samples from the conditional distribution of $\vec{X}$ given that it lies in $F_i$, which is much more challenging. This is achieved by applying Markov Chain Monte Carlo (MCMC) methods. In the original SuS work a component-wise modified Metropolis-Hasting (MMH) algorithm \cite{Au2001} was introduced. In the subsequent years, many new MCMC approaches have been devised because the efficiency and accuracy of SuS can strongly depend on the efficacy of the MCMC sampler to identify independent samples in each conditional level. These include algorithms based on repeated generations of the candidate state until the first acceptance criterion is satisfied \cite{santoso2011modified}, the related delayed rejection method \cite{miao2011modified, au2012discussion, zuev2012bayesian}, methods that translate gradually samples from the prior to samples from the posterior distribution through a combination of importance sampling and MCMC \cite{Ching2007} and methods to adaptively update the proposal density in each conditional level \cite{Iason2015}. Recently, \cite{shields2021subset} Shields et al. proposed to use the affine invariant ensemble sampler with stretch moves \cite{Goodman2010}, to perform SuS for problems where distributions or conditional distributions are strongly non-Gaussian, highly dependent, and/or degenerate. This is the method employed herein.

\section{Multimodel Inference}
\label{InfomModel}

Multimodel inference, presented in the pioneering work of Burnham and Anderson \cite{burnham2004multimodel}, is an approach for quantifying model-form and parametric uncertainty from small data sets. Unlike classical inference for model selection, where the goal is to identify a single ``best'' model, the multimodel inference process argues that it is often impossible to uniquely identify a single best model, especially when data are sparse. Instead, multimodel inference aims to retain a set of models that are potentially representative of the data and, moreover, to assign to each model a probability that it is the ``best'' model. This multimodel inference process can be formalized in either a Bayesian \cite{Zhang2018a} or and information theoretic setting \cite{burnham2004multimodel, Zhang2018}. In this work, we will use the information theoretic setting simply because it does not require us to undertake the difficult task of computing model evidence. However, we emphasize that the Bayesian framework is equally applicable and does not fundamentally change the proposed reliability approach.

\subsection{Information Theoretic Multimodel Selection}

\noindent

Consider that we have a limited data set, denoted $\textbf{d}$, of randomly drawn samples of a random variable $X$ having unknown probability distribution. Next, consider a set of candidate probability models $\{\mathcal{M}_l\}_{l=1}^{N_d}$ that represents our initial belief about the distribution of $X$. In an information theoretic framework, the suitability of each model to represent the data can be assessed by measuring the Kullback-Leibler (K-L) information loss \cite{Kullbeck1951} associated with each model. A robust way to estimate the K-L information loss is through the so-called Akaike Information Criterion (AIC) \cite{Akaike1974}
associated with model $\mathcal{M}_l)$ defined as
\begin{equation}\label{AIC}
\mbox{AIC}^{(l)} = -2\log(\mathcal{L}(\boldsymbol{\theta}_l^\star|\textbf{d}, \mathcal{M}_j)) + 2k_l
\end{equation}
where $\mathcal{L}(\cdot)$ is the likelihood function for model $\mathcal{M}_l$ evaluated at the maximum likelihood parameters $\boldsymbol{\theta}_l^\star$ given data $\textbf{d}$ and  $2k_l$ is a bias correction factor where $k_l$ is the dimension of the parameter vector $\boldsymbol{\theta}_l$.  Since the AIC is an asymptotic quantity relying on large data sets, an extension of the AIC introduced in \cite{Hurvich1989}, is recommended when dealing with small data sets. The modified criterion, given by
\begin{equation}
\mbox{AIC}_c^{(l)} = -2\log(\mathcal{L}(\boldsymbol{\theta}^\star|\textbf{d}, \mathcal{M}_l)) + 2k_l +  \frac{(2k_l^2+k_l)}{n-k_l-1}
\end{equation}\label{AIC_c}
incorporates a second-order bias correction term where $n$ is the size of the data set $\textbf{d}$.  

Using the $\mbox{AIC}_c$ estimates of the information loss for multiple model, it is possible to estimate probabilities that each model is the best model. To estimate model probabilities, we first need to establish a relative scale for the $\mbox{AIC}_c$. This is achieved by defining the ``relative'' information loss as $\Delta_A^{(l)} = \mbox{AIC}_c^{(l)} -\min(\mbox{AIC}_c^{(1)}, \ldots, \mbox{AIC}_c^{(N_d)})$ for each model $\mathcal{M}_l$ such that the model having the lowest information loss has $\Delta_A^{(l)}=0$. Model probabilities can be calculated from the $\mbox{AIC}_c$ as
\begin{equation}
{\pi}_l = \mathbb{P}(\mathcal{M}_l|\textbf{d}) = \frac{\exp{\bigg(\frac{-\Delta_A^{(l)}}{2}\bigg)}}{\sum_{l=1}^{N_d} \exp{\bigg(\frac{-\Delta_A^{(l)}}{2}\bigg)}}\label{ACC_pi}
\end{equation}
The interpretation of the ratio in Eq.\ \eqref{ACC_pi} as a probability is justified through Bayesian arguments using a class of ``savvy'' priors (see \cite{burnham2004multimodel} for justification).

\subsection{Model Parameter Uncertainty}

The model probabilities in Eq.\ \eqref{ACC_pi} provide an estimate of model-form uncertainty. However, for each model we also need to quantify uncertainties associated with the parameters $\boldsymbol{\theta}_l$ $l=1, \ldots, N_d$, where $N_d$ is now the number of models from the original model set having non-negligible probability. Here, we apply a standard Bayesian framework for parameter estimation. We begin by assigning 
a prior pdf $p(\boldsymbol{\theta}_r;\mathcal{M}_l)$ to the model parameters that reflects our current belief about the parameter values. Appropriate prior selection is crucial when dataset are small \cite{Zhang2018a}. A simple choice is the uniform prior $p(\boldsymbol{\theta}_l;\mathcal{M}_l)=1/N_d$ that weights all parameters equally within a specific range, which will be used herein for convenience in methodology demonstration. However, more advanced techniques relying on hyper-priors \cite{bangalore2020multimodel} may be appropriate to more rigorously account for uncertainty in prior probabilities. 

The posterior pdf $p^\star(\boldsymbol{\theta}_l|\vec{d};\mathcal{M}_l)$ given $\textbf{d}$ that represents our updated belief is estimated through Bayes' Rule as
\begin{equation}
p^\star(\boldsymbol{\theta}_l|\textbf{d}, \mathcal{M}_l) = \frac{p(\textbf{d}|\boldsymbol{\theta}_l, \mathcal{M}_l)p(\boldsymbol{\theta}_l;\mathcal{M}_l)}{p(\textbf{d};\mathcal{M}_l)} \propto \mathcal{L}(\boldsymbol{\theta}_l|\textbf{d}, \mathcal{M}_l)p(\boldsymbol{\theta}_l;\mathcal{M}_l)
\label{Bayes}
\end{equation}
where $\mathcal{L}(\boldsymbol{\theta}_l|\textbf{d}; \mathcal{M}_l)=p(\textbf{d}|\boldsymbol{\theta}_l;\mathcal{M}_l)$ is the likelihood of the parameters $\boldsymbol{\theta}_l$ given the observed data $\textbf{d}$,  and $p(\textbf{d};\mathcal{M}_l)$ is the  evidence estimated by
\begin{equation}
p(\textbf{d};\mathcal{M}_l) = \int p(\textbf{d}|\boldsymbol{\theta}_l;\mathcal{M}_l)p(\boldsymbol{\theta}_l;\mathcal{M}_l)d\boldsymbol{\theta}_l
\end{equation}
Computing the evidence is not trivial and therefore, a Markov Chain Monte Carlo (MCMC) approach is applied to draw samples from the posterior $p^\star(\boldsymbol{\theta}_l|\textbf{d}; \mathcal{M}_l)$.

\subsection{Obtaining a Set of Candidate Models}

This multi-model selection process allows us to identify a set of models $\{\mathcal{M}_l\}_{l=1}^{N_d}$ with associated probabilitites $\pi_l$ and the joint posterior pdf of each model's parameters. In theory, this set comprises an infinite number of probability distributions (a finite number of probability models with continuous posterior parameter joint pdfs). In practice, Bayes' Rule is applied for parameter estimation to each of the $N_d$ plausible models $\mathcal{M}_l$ to sample a sufficiently large and statistically representative set (of size $n_\theta$) of model parameters $\boldsymbol{\theta}_l$ using MCMC. The result is an aggregate set of $N_d \times n_\theta$ candidate distributions from which the data may have been drawn.

When conducting reliability analysis, one may choose to retain the full set of $N_d \times n_\theta$ distributions and their associated model probabilities. For simplicity, we perform Monte Carlo sampling from this set to obtain a set of $n_c$ equally weighted probability distributions denoted $M_j,$ $j=1,\dots,n_c$. That is, we randomly select $n_c$ candidate pdfs to represent the data. This is done by first randomly selecting a probability model according to the model probabilities $\pi_l$ and then randomly drawing the model parameters from the joint posterior density obtained from MCMC. As we will see, $n_c$ can be made arbitrarily large with only minor influence on computational expense. 

\section{Imprecise subset simulation}

\noindent
Subset simulation is efficient and precise when probability models for the input random variables (and hence the conditional levels) are uniquely prescribed. However, when probability models are uncertain and represented by multiple models, special treatment is needed to estimate probability of failure using SuS. For this, we can discuss two possible
approaches. 

\vspace{6pt}
\noindent
{\it Approach 1: Constant conditional probabilities and varying conditional levels.} In this approach, we allow for unique conditional levels to be defined for each set of probability models such that the conditional probabilities are fixed. This is consistent with the conventional SuS approach, where each conditional level is defined according to a prescribed conditional probability level. 

\begin{figure}[!ht]
	\centering
	\includegraphics[width=0.7\textwidth]{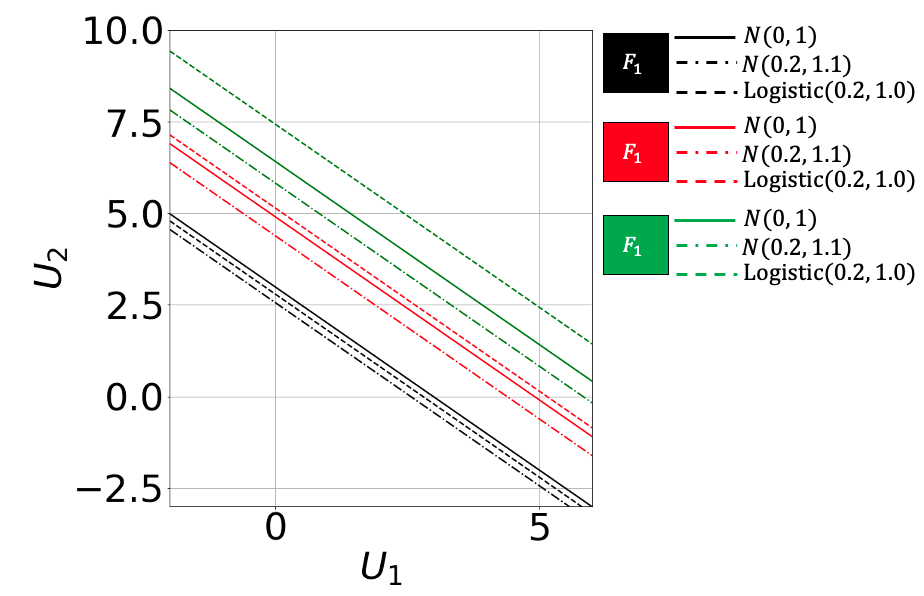}
	\caption{For distribution models  $M_1=N(0,1)$, $M_2=N(0.2, 1.1)$, and $M_3=\text{Logistic}(0.2, 1)$, running SuS will result in  different conditional levels $F_i(M_j)$, for $i, j=1, \ldots, 3$, respectively, if the conditional probability $P_i$ is held constant.}
	\label{fig:approach1}
\end{figure}

To illustrate this approach, consider a simple reliability problem with two iid random variables $\textbf{U}=[U_1, U_2]$ having a linear performance function given by $g(\textbf{U}) = \beta \sqrt{2} - \sum_{i=1}^2 U_i$, where $\beta$ is a specifiied reliability index. Let us now assume that, through multimodel inference, we identify the \textit{Normal} and \textit{Logistic} distributions as candidate models to represent our data. Let us further assume that three of the $n_c$ candidate probability models are the following: $N(0,1)$, $N(0.2, 1.1)$ and $Logistic(0.2, 1)$. The conditional levels for SuS with random variables having these  probability models are depicted in Fig.\ref{fig:approach1} where each conditional level $F_i$  $i=1, \ldots, 3$ (represented with a different colour) corresponds to a set of three lines, one for each model. Note that each conditional level is the average conditional level from 100 independent SuS runs with fixed conditional probabilities $P_i=0.1$. Although these probability models are quite similar, the conditional levels vary significantly and introduce significant uncertainty in the reliability estimates.   

From Figure \ref{fig:approach1} can see that, because the conditional levels move with each change of the probability models, we cannot estimate probability of failure from a single set of SuS samples. Hence, a unique subset simulation is required to solve for $P_F$ for each set of distributions independently. Given that the multi-model approach identifies potentially thousands of sets of candidate distributions, this is clearly not
tractable, especially for expensive performance models.

\vspace{6pt}
\noindent
{\it Approach 2: Constant conditional levels and varying conditional probabilities.} An alternate approach is to fix the conditional levels and allow the conditional probabilities to vary for different sets of candidate probability models. Although different from the conventional SuS approach, this approach has the advantage that, because the conditional levels do not change, we can potentially leverage a single set of SuS samples to estimate failure probabilities for different candidate probability models. Practically, this means conducting only a single SuS and applying a reweighting scheme to modify the conditional probabilities for different probability models. 

The challenge with this approach is to identify the appropriate conditional levels. Conventional SuS defines the conditional levels indirectly through the assignment of target conditional probabilities. However, since this approach considers the conditional probabilities to change for each set probability models, we need an alternate strategy to identify the conditional levels. In the proposed approach below, we do so by assigning the conditional probabilities from a single, optimal sampling density. This process is described next.  

\subsection{Proposed Approach}
Here, we propose a method that follows {\it Approach 2} to leverage a single SuS to estimate probabilities of failures from multimodel sets. That is, we fix the conditional performance levels, $F_i$, and allow the conditional probabilities to vary with each probability distribution in the multi-model set, i.e. $P_i\to P_{ij} \equiv P_i(M_j)$, $j=1,\dots, n_c$ where $M_j$ corresponds to a probability distribution (model form and parameters) drawn from the set of $n_c$ candidate distributions. To identify the conditional levels, we first define an optimal sampling density. According to the \cite{Zhang2018}, the optimal sampling density can be identified by minimizing the total expected mean square difference
\begin{equation}
\begin{aligned}
& \underset{q}{\text{minimize}}
& &\hat{\mathcal{T}}[q]=E_{\theta}\left [ \int_{\Omega }{\hat{F}(\bm{x}, \bm{\theta}, q(\bm{x}))}d\bm{x} \right] \\
& \text{subject to}
& &\hat{\mathcal{I}}[q] =  \int_{\Omega}{q(\bm{x})d\bm{x}}-1=0 \label{opt_EMSD}
\end{aligned}
\end{equation}
where $\Omega$ is the sample space, and
\begin{equation}
{\hat{F}(\bm{x}, \bm{\theta}, q(\bm{x}))}={  \frac { 1 }{ 2 } \sum_{l=1}^{N_d}{ \left( { p_l(\bm{x} | \bm{\theta}) } - { q(\bm{x}) }  \right) ^{ 2 }}   } \label{eq:MSD_funcitonal} 
\end{equation}
This minimization is shown to have a closed form solution \cite{Zhang2018} corresponding to the weighted mixture distribution given by:
\begin{equation}
\hat{q}^{*}(\bm{x}) =  \sum_{l=1}^{m}\pi_l E_{\theta} \left [ { p_l(\bm{x}|\bm{\theta})} \right]  \label{eq: opt_MSD3}
\end{equation}
where $\pi_l$ is the $\textup{AIC}_c$ model probability (see Eq. (\ref{ACC_pi})) for model $\mathcal{M}_l$ satisfying $\sum_{l=1}^{N_d}\pi_l=1$.

We then perform a ``baseline'' SuS using this optimal sampling density having specified conditional probabilities $P_i^{opt}$ for each conditional level $i$ (typically $P_i^{opt}=0.1$, $\forall i$). From this baseline SuS, we retain all samples and the associated conditional levels. We then reweight the samples according to each of the $n_c$ candidate probability models using importance sampling. 
Recall that the importance sampling estimator for a function $f(X)$ of a random variable $X$ having pdf $p(x)$, with sampling drawn from the importance sampling density $q(x)$ can be expressed as:
\begin{equation}
    E_p[f(x)] = E_q[f(X)] = \dfrac{1}{N} \sum_{j=1}^N \dfrac{p(x_j)}{q(x_j)}f(x_j)
\end{equation}
where $N$ is the number of samples. For a single prescribed probability model at conditional level $i$, $f(X)$ is defined as the indicator function $I_{F_{i+1}}=I(X\in F_{i+1})$. This expectation can be used to compute the conditional probability as
\begin{equation}
    P_{i+1} = P(F_{i+1}|F_i) \approx \dfrac{1}{N_i} \sum_{k=1}^{N_i} \dfrac{p(x_k)}{q(x_k)}I_{F_{i+1}}(x_k)
    \label{eqn:IS_cond_P}
\end{equation}
where $N_i$ is the number of samples drawn in conditional level $i$. The probability of failure is then computed as 
\begin{equation}
    P_F = \prod_{i=1}^{m} P_i
\end{equation}
where $m$ is the number of conditional levels.

This process is repeated for each candidate probability model $\mathcal{M}_l$, where $p(x)$ in Eq.\ \eqref{eqn:IS_cond_P} is replaced with the probability density $p_j(x)$ associated with model $M_j$, $j=1,\dots,n_c$ and $q(x)$ is the optimal sampling density, $\hat{q}^{*}(\bm{x})$. The corresponding probability of failure is then estimated as:
\begin{equation}\label{eq:Failure_prob_imprecise}
P_{Fj} = P_F(M_j) = \prod_{i=1}^{m} P_{i}(M_j) = \prod_{i=1}^{m} P_{ij}
\end{equation}
We then compile the statistical set of failure probabilities $P_{Fj}$. Recall that we draw $M_j$ in such a way that all $M_j$ have equal probability of occurrence. From these values, we can construct the empirical cumulative distribution of the failure probability given the multimodel set, which provides a measure of the uncertainty in $P_F$ associated with probability model form uncertainty. Moreover, we emphasize that the computational cost of the proposed method depends very little on the size of $n_c$. An increase in $n_c$ does not require any additional performance function evaluations. It simply corresponds to an increased number of sample re-weightings.

We next elaborate the specific steps of the proposed methodology.


\subsection{Steps of the proposed method}

\noindent
{\it Step 1: Multimodel Inference}

\noindent
For each random variable $X_{\alpha}$, $\alpha=1, \ldots, n$
\begin{itemize}
	\item[1.] Identify the set of candidate probability models and their associated model probabilities $\{\mathcal{M}_r, \pi_l\}_{l=1}^{N_d}$ using information theoretic (or Bayesian, \cite{Zhang2018a}) multi-model selection as described in Section 3.
	
	\item[2.] For each model $\mathcal{M}_l$: i) identify the joint posterior parameter probability density $p(\boldsymbol{\theta}_l|\vec{d};\mathcal{M}_l)$ using Bayesian inference  and, ii) sample ($n_{\theta}>10,000$) from this density  using MCMC.

	\item[3.] Randomly select $n_c$ (large) models from the  pool of $N_d\times n_{\theta}$ candidate models. 

\end{itemize}
The result is a Monte Carlo set of $n_c$ candidate probability models. We aim to estimate the failure probability of the system for each of these candidate models.

\vspace{6pt}
\noindent
{\it Step 2: Optimal Sampling Density}

\noindent
Construct the optimal marginal sampling density, $\hat{q}^{*}(\bm{x})$, for each random variable according to Eq.\ \eqref{eq: opt_MSD3}. Assuming independent random variables, construct the joint probability density function as 
\begin{equation} \label{joint_opti_density_i}
q_n^\star(\vec{x})=\prod_{\alpha=1}^n \hat{q}_\alpha^\star(\vec{x})
\end{equation}

\vspace{6pt}
\noindent
{\it Step 3: Subset Simulation}

\noindent
Run a single (optimal) SuS 
using the optimal sampling density to identify the conditional performance levels $F_i^{opt}$ and obtain samples $\vec{x}_{i}^{opt}$ in each conditional level.  

\vspace{6pt}
\noindent
{\it Step 4: Importance Sampling Re-weighting}

\noindent
Apply importance sampling (IS) to re-weight the conditional probabilities $P_i$ for each distribution model $M_j$. More specifically, at subset $i$, for each model  we calculate the importance weight $w_{j}$ at each sample point $\vec{x}_{k}^{opt}$ as 
\begin{equation}
w_{j}(\vec{x}_{k}^{opt})=\frac{p_j(\vec{x}_{k}^{opt})}{q^\star_n(\vec{x}_{k}^{opt})}, \quad j=1,\ldots,n_c
\end{equation}
where $p_j(\cdot) = p(\boldsymbol{\theta}_{j}|\textbf{d}, M_j)$. The corresponding conditional probability, $P_{ij}$ for model  $M_j$ is estimated  as
\begin{equation}
P_{ij}=\frac{1}{N_i}\sum_{k=1}^{N_i} w_j(\vec{x}_k^{opt})\text{I}_{F_i^{opt}}(\vec{x}_k^{opt})
\end{equation}
and the probability of failure for model $\mathcal{M}_{t}$ is estimated as
\begin{equation}
	P_{Fj} = \prod_{i=1}^m P_{ij}
\end{equation}
	
This step is repeated $n_c$ times; once for each of the candidate probability models. The result is a stochastic set of failure probabilities $\{P_{Fj}\}_{j=1}^{n_c}$ that incorporate the effects of model form and parameter uncertainty on the estimated probability of failure.

\section{Numerical examples}

\noindent
We apply the proposed imprecise SuS approach to quantify uncertainty in structural reliability estimates for two structural problems: i)  a simply supported rectangular plate under uniaxial compression and, ii)  a two-story linear structure subjected to a narrow-banded ground acceleration. In both examples, we consider data set of increasing size (25, 100, 500, and 1000) for each random variable obtained by randomly sampling from a specified ``true'' distribution. Our set of candidate probability models consists of seven different families: \textit{lognormal,  gamma, logistic, inverse Gaussian, Maxwell, Levy} and \textit{normal}. 

Subset simulation,  MCMC samplers and the multi-model selection framework used in this work are all available in the open-source UQpy Python toolbox (\cite{olivierUQpy, UQpy}) and all calculations presented herein have been performed using this software.

\subsection{Example 1}

\noindent
In this example we study failure of a simply supported rectangular plate under uniaxial compression given uncertainties in the material properties. The normalized buckling strength for an imperfect plate is given in a closed form by \cite{faulkner1975review,carlsen1977simplified} 
\begin{equation}\label{psi}
\psi = \bigg(\frac{2.1}{\lambda}-\frac{0.9}{\lambda^2}\bigg)\bigg(1-\frac{0.75\delta_0}{\lambda}\bigg)\bigg(1-\frac{2\eta t}{b}\bigg)
\end{equation}
where $\lambda=\frac{b}{t}\sqrt{\frac{\sigma_0}{E}}$,  $b$ is the plate width, $t$ is the plate thickness, $\sigma_0$ is the yield stress, $E$ is the elastic modulus, $\delta_0$ is the initial out-of-plane deflection, and $\eta$ is the residual stress caused by welding at the boundaries. Nominal values and statistics for these six variables, taken from \cite{Hess2002, soares1988uncertainty} are provided in Table \ref{tab:my_label}. 
\begin{table}[!ht]
\resizebox{\textwidth}{!}{
	\centering
	\begin{tabular}{ccccc}\toprule
		{Variable} & {Physical meaning} & {Nominal value} & {Mean} & {COV}  \\ \midrule
		$b$  & Width (in) & 24 & $0.992\times 24$ & 0.028 \\
		$t$  & Thickness (in)  & 0.5 & $1.05\times0.5$ & 0.044 \\
		$\sigma_0$  & Yield strength (ksi)  & 34 & $1.3\times34$  & 0.1235 \\
		$E$  & Elastic modulus  & 29000 (ksi) & $0.987\times29000$  & 0.076      \\ 
		$\delta_0$  & Initial deflection (in)  & 0.35 & $1.0\times0.35$   & 0.05\\
		$\eta$  & Residual stress parameter (dimensionless)  & 5.25 & $1.0\times5.25$  & 0.07 \\ \bottomrule
	\end{tabular}}
	\caption{Statistical properties of plate material, geometry and imperfection parameters.}
	\label{tab:my_label}
\end{table}

A sensitivity analysis performed in \cite{Zhang2018} showed that although all six variables affect the buckling strength, the material parameters (elastic modulus and yield strength), account for $\sim70$\% of the total variability in plate strength according. We will therefore include only these two variables in our reliability analysis. We express the yield strength as $\sigma_0= \hat{\sigma_0} +34$ where the ``true'' distribution of the $\hat{\sigma_0}$ is lognormal  with mean $\mu_{\hat{\sigma_0}}=10.2$ and standard deviation $\sigma_{\hat{\sigma_0}}=5.4587$. The ``true'' distribution of the elastic modulus $E$ is  selected  to be normal with mean and coefficient of variation given in Table \ref{tab:my_label} \cite{Hess2002}. All other values are fixed at their nominal value.

\begin{figure}[!ht]
	\centering
	\begin{subfigure}[t]{0.47\textwidth}
		\centering
		\includegraphics[width=\textwidth]{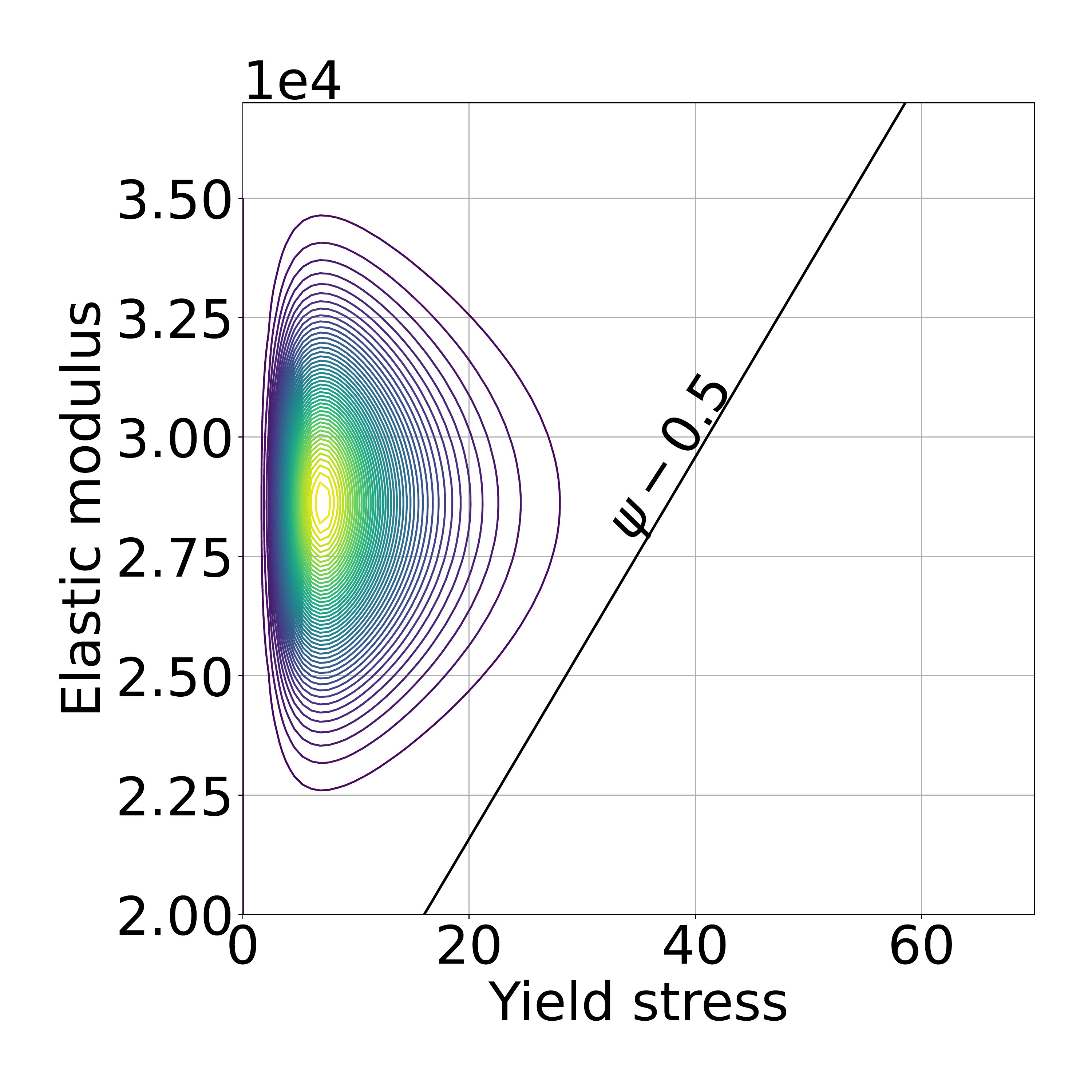}
		\caption{}
	\end{subfigure}
	\begin{subfigure}[t]{0.47\textwidth}
		\centering
		\includegraphics[width=\textwidth]{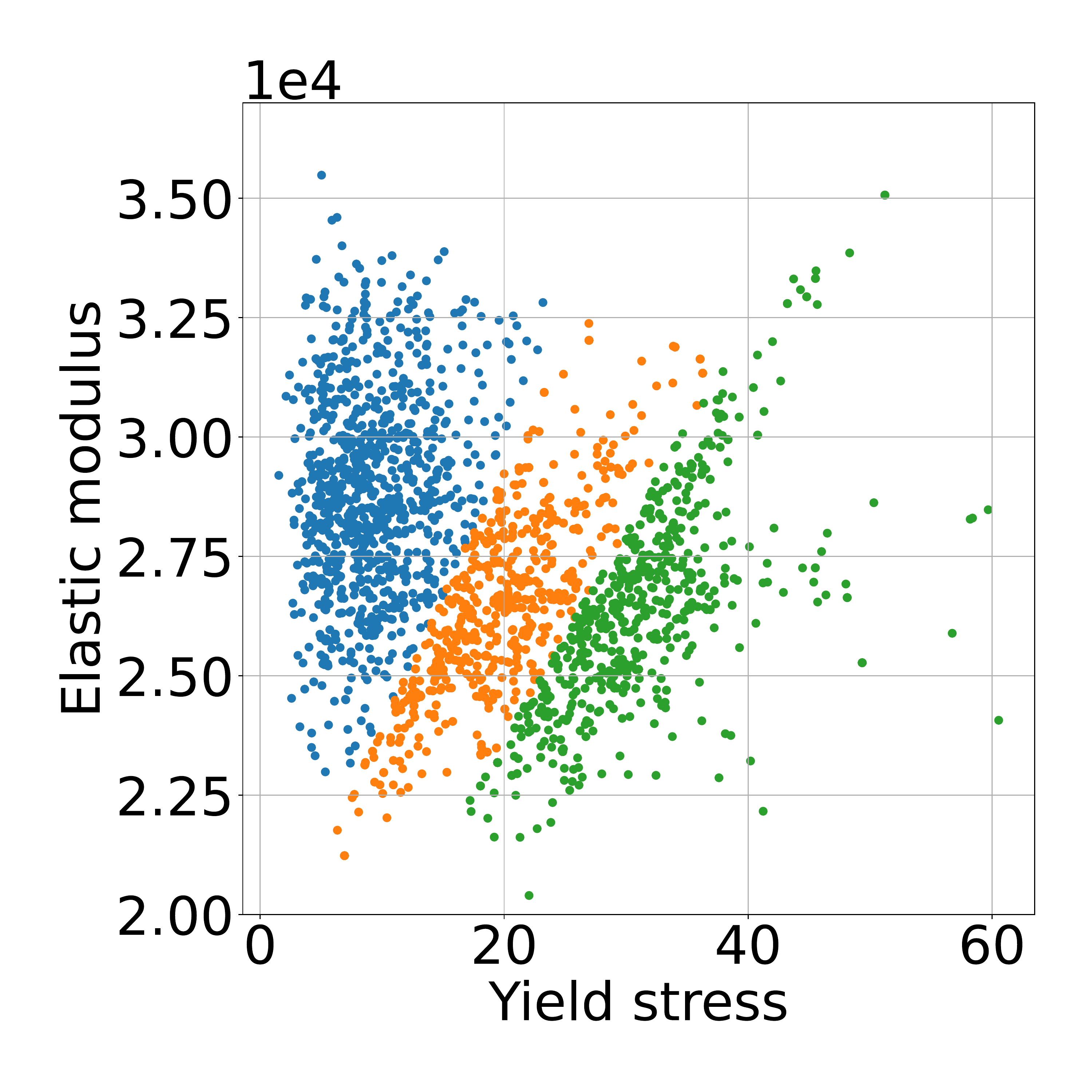}
		\caption{}
		\end{subfigure}
	\caption{Example 1: (a) Limit-state function in the original space of material parameters, (b) Samples in each conditional leve from SuS.}
	\label{fig:SuS_ex2}
\end{figure}

We define probability of failure as $P_F = P(\psi <0.5)$ and determine the ``exact'' probability of failure as $P_F=0.003$ by Monte Carlo simulation with $10^6$ samples.  Using SuS with 1000 samples per subset and conditional probability $P_i=0.1$, the probability of failure is found to be $P_F=0.0031$ with a coefficient of variation (c.o.v) equal to 0.47, verifying that SuS provides an accurate reliability estimate. Figure \ref{fig:SuS_ex2}(a)  depicts the limit-state function in the original space of the material parameters, while Figure \ref{fig:SuS_ex2}(b) shows the samples in each conditional level from a single subset simulation.

\subsection*{Imprecise Subset Simulation}
\begin{figure}[!ht]
	\centering
	\begin{subfigure}[t]{0.40\textwidth}
		\centering
		\includegraphics[width=\textwidth]{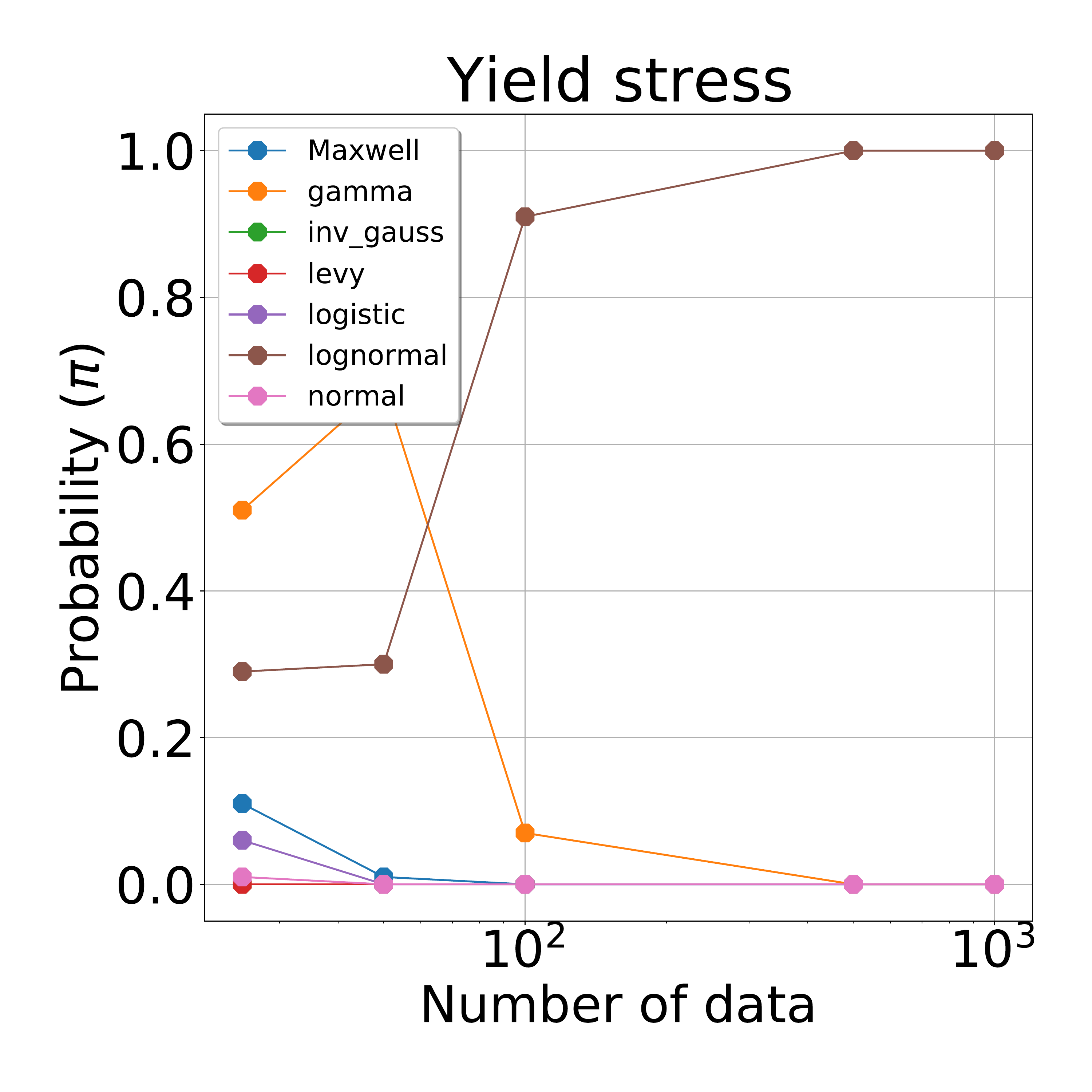}
	\end{subfigure}
	\begin{subfigure}[t]{0.40\textwidth}
		\centering
		\includegraphics[width=\textwidth]{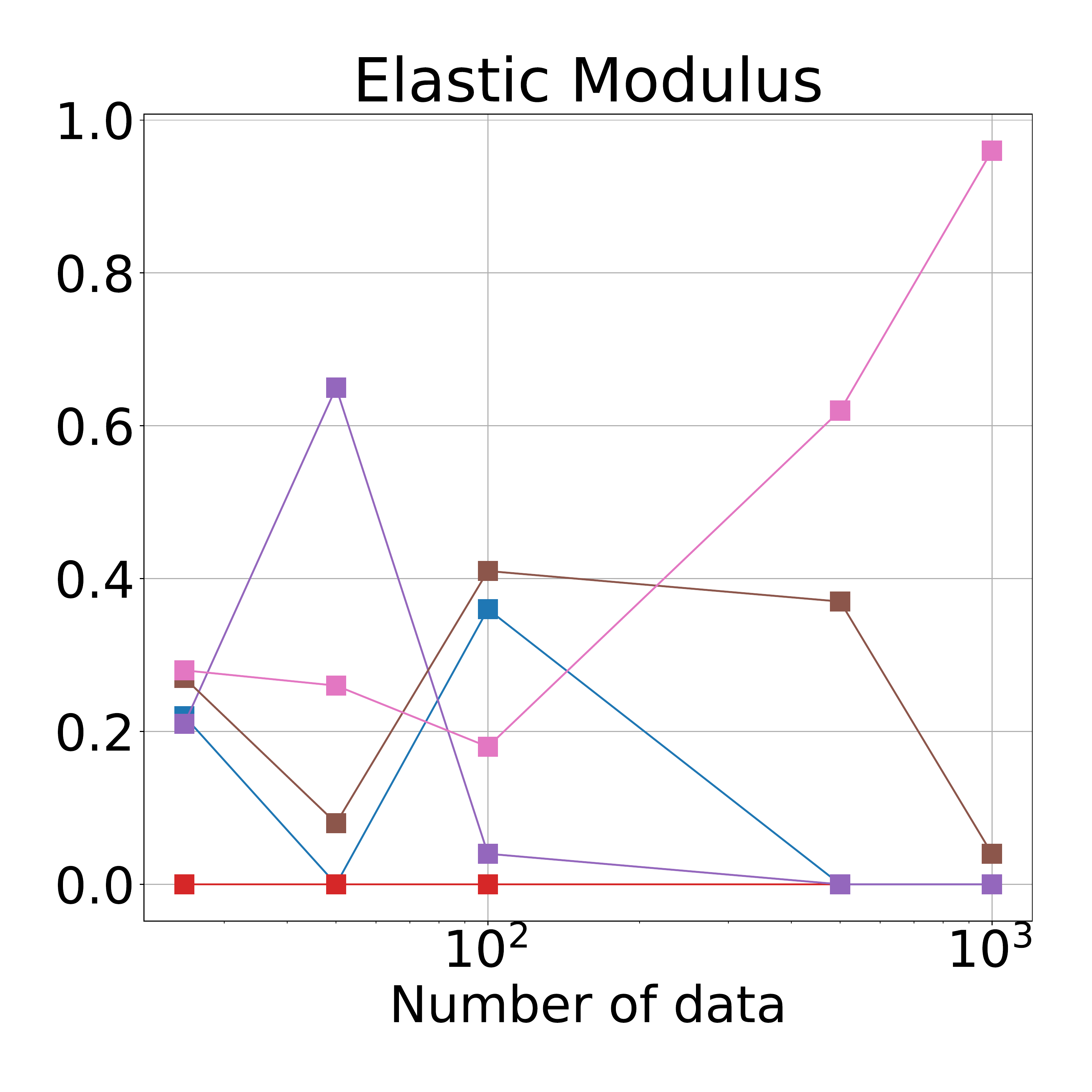}
	\end{subfigure}
	\caption{Example 2: Model probabilities as a function of data set size.}
	\label{fig:Prob_ex2}
\end{figure}

\noindent
Next, we consider uncertainty in the distributions of the material parameters due to limited data and apply the proposed approach for imprecise SuS. We consider cases of varying data set sizes from 25 to 1000 measurements. Figure \ref{fig:Prob_ex2} plots the model probabilities for each model (i.e.\ the probability of that modeling being  the ``best'' distribution model) as a function of data set size. From this figure it is clear that   the  true distribution family for  yield stress (which is the \textit{lognormal}) can be identified with $85\%$ certainty with  just 100  yield stress measurements. On the other hand,  even a large dataset with more than 500 measurments is not sufficient to identify the true distribution model  for the elastic modulus with certainty.  In the latter case, we need 1000 measurements in order to be confident that the true distribution comes from the \textit{Normal} distribution family.

\begin{figure}[!ht]
	\centering
	\begin{subfigure}[t]{0.40\textwidth}
		\centering
		\includegraphics[width=\textwidth]{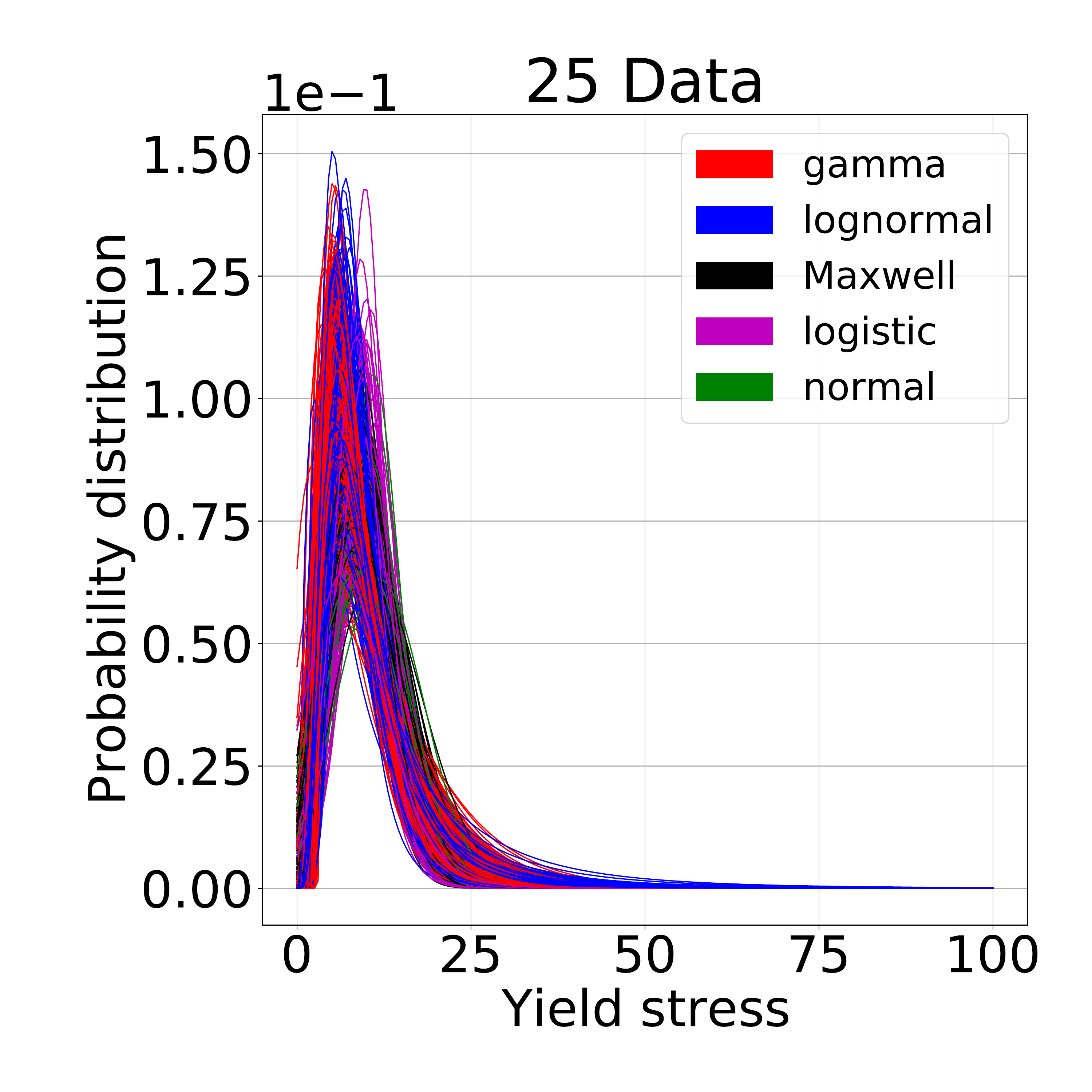}
	\end{subfigure}
	\begin{subfigure}[t]{0.40\textwidth}
		\centering
		\includegraphics[width=\textwidth]{Example2_Distributions_Yieldstress_25.pdf}
	\end{subfigure}
	
	\begin{subfigure}[t]{0.40\textwidth}
		\centering
		\includegraphics[width=\textwidth]{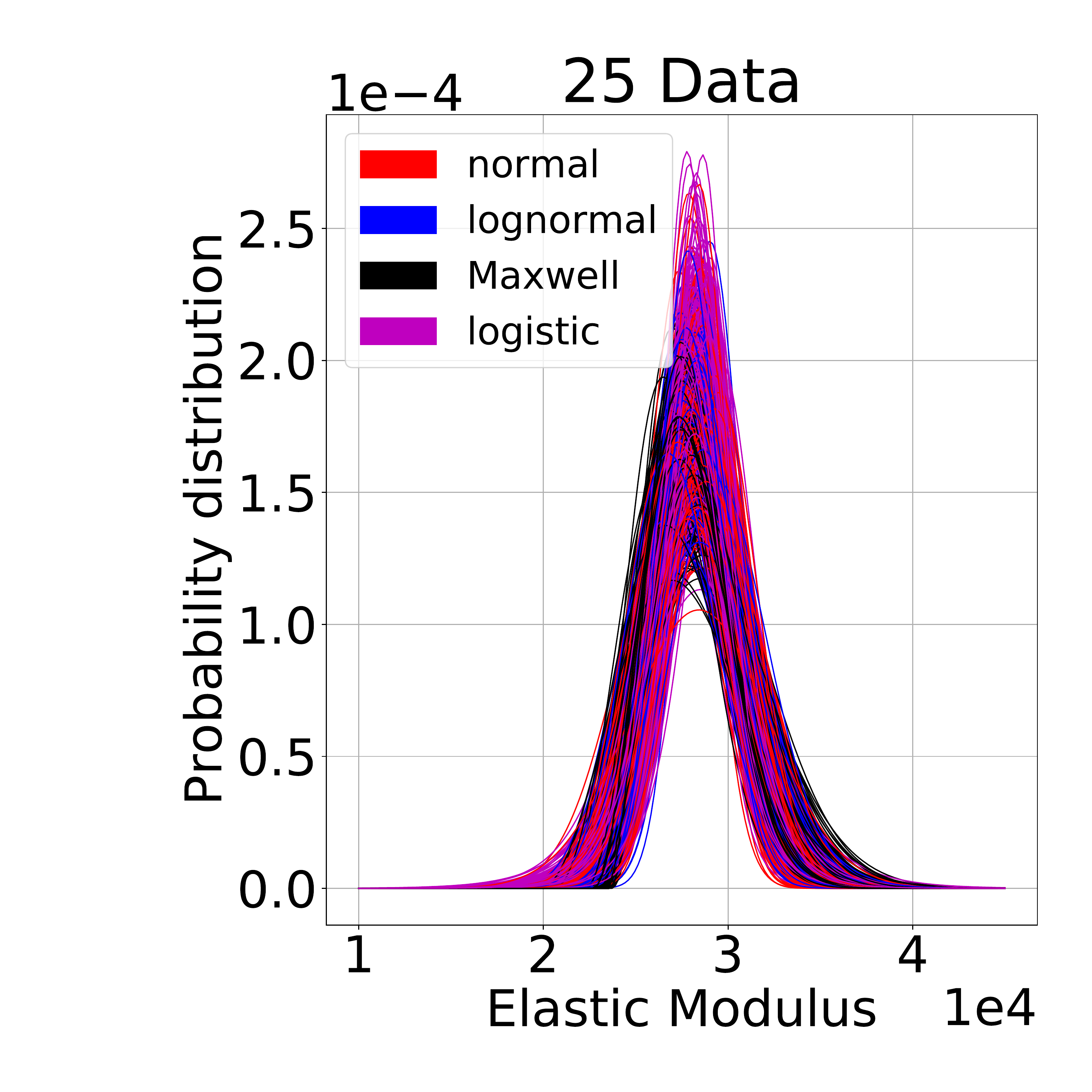}
	\end{subfigure}
	\begin{subfigure}[t]{0.40\textwidth}
		\centering
		\includegraphics[width=\textwidth]{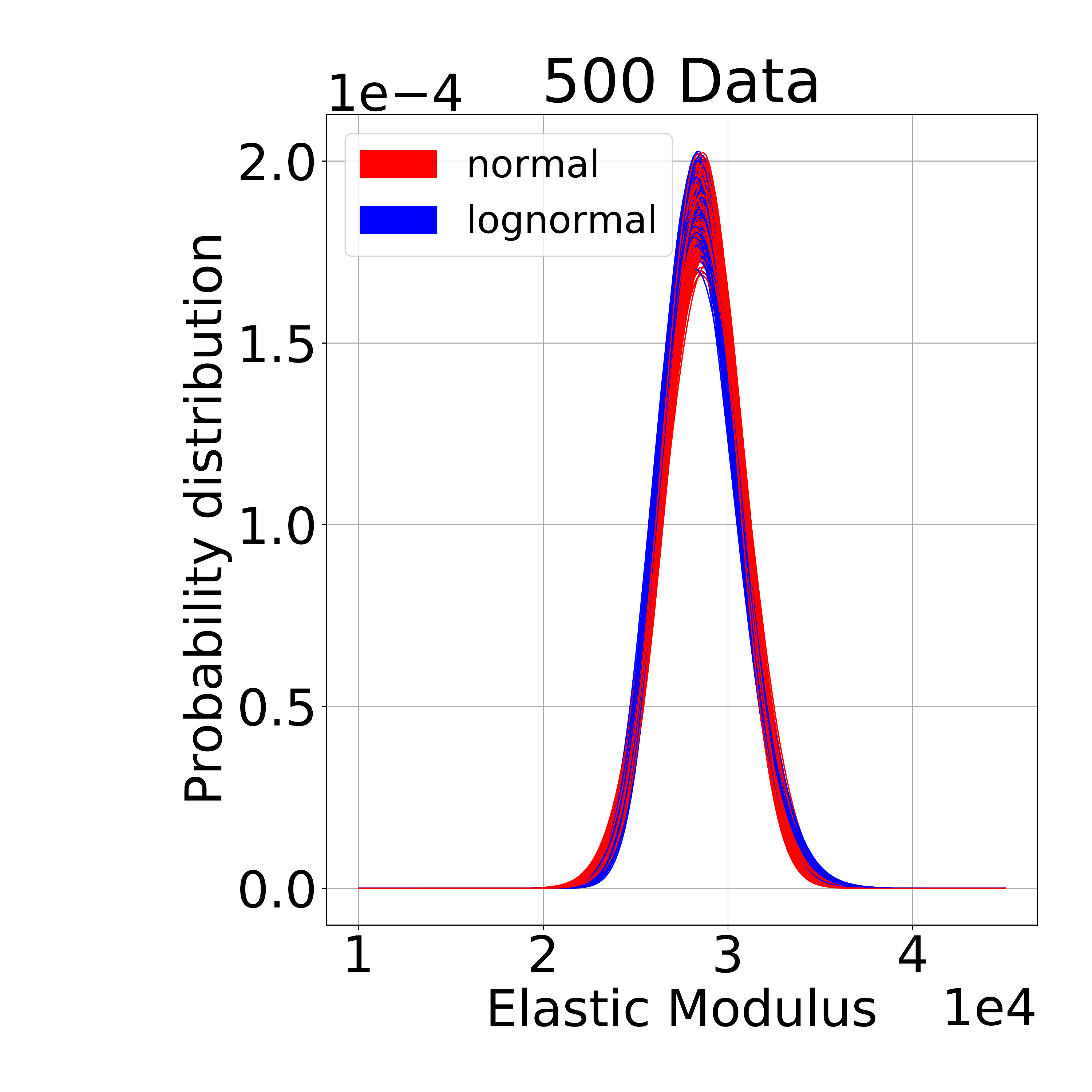}
	\end{subfigure}
	\caption{Example 2: Sample sets of probability distributions from multimodel selection given 25 and 500 data for the shifted yield stress, $\hat{\sigma}_0$, (top) and the elastic modulus (bottom).}
	\label{fig:Dists_ex2}
\end{figure}

Figure \ref{fig:Dists_ex2} shows a  sampled set of $n_c=1000$ candidate distributions for the yield stress and the elastic modulus for 25 and 500 available measurements.  The total uncertainty in the material parameter distributions arising from data sparsity is reflected on the clouds of distributions that narrow around the true distribution as the data set size grows. The distributions are colored according to their distribution family and we see that, again as data set size increases, we narrow to only a few candidate distribution types.  

\begin{figure}[!ht]
	\centering
	\begin{subfigure}[t]{0.40\textwidth}
		\centering
		\includegraphics[width=\textwidth]{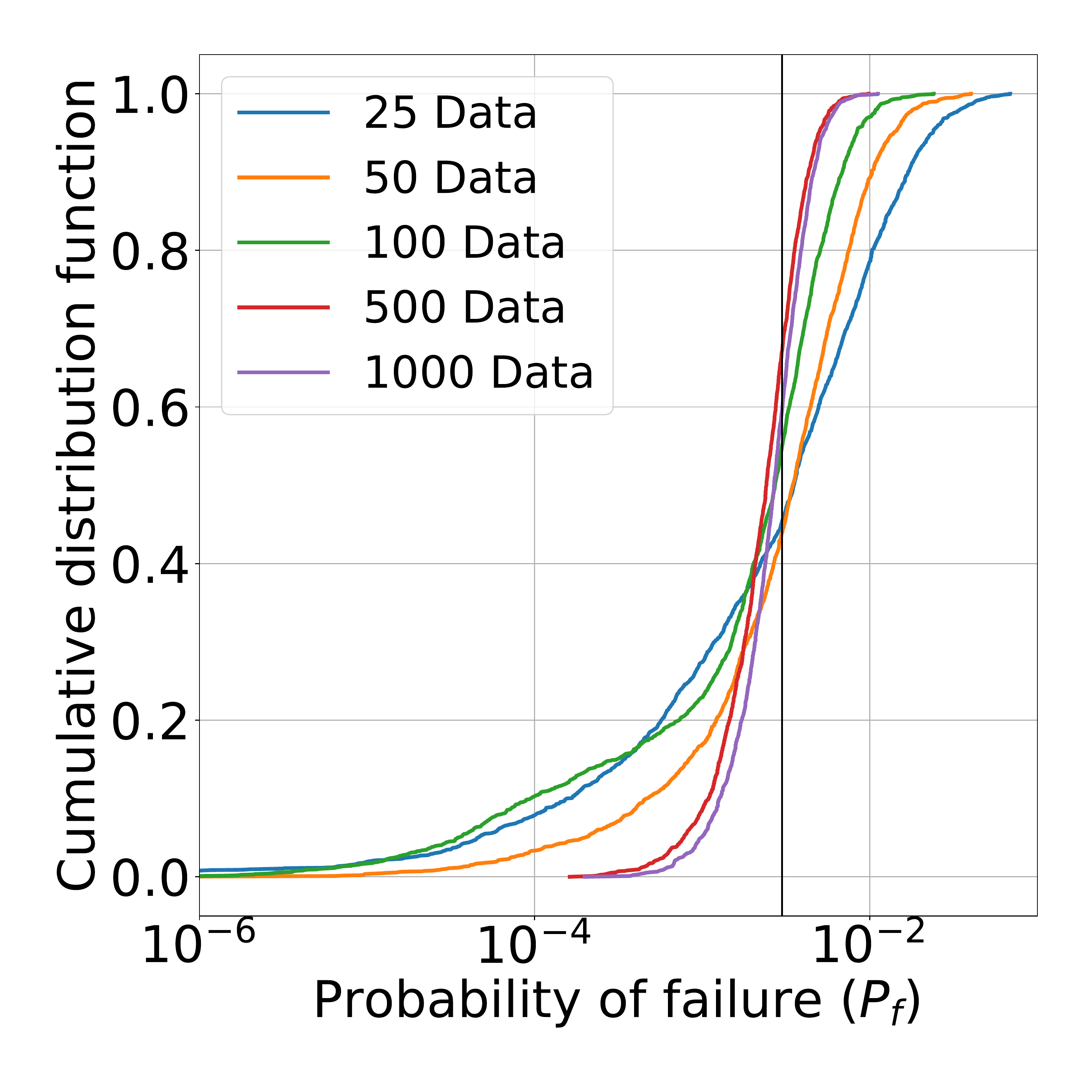}
		\caption{}
	\end{subfigure}
	\begin{subfigure}[t]{0.40\textwidth}
		\centering
		\includegraphics[width=\textwidth]{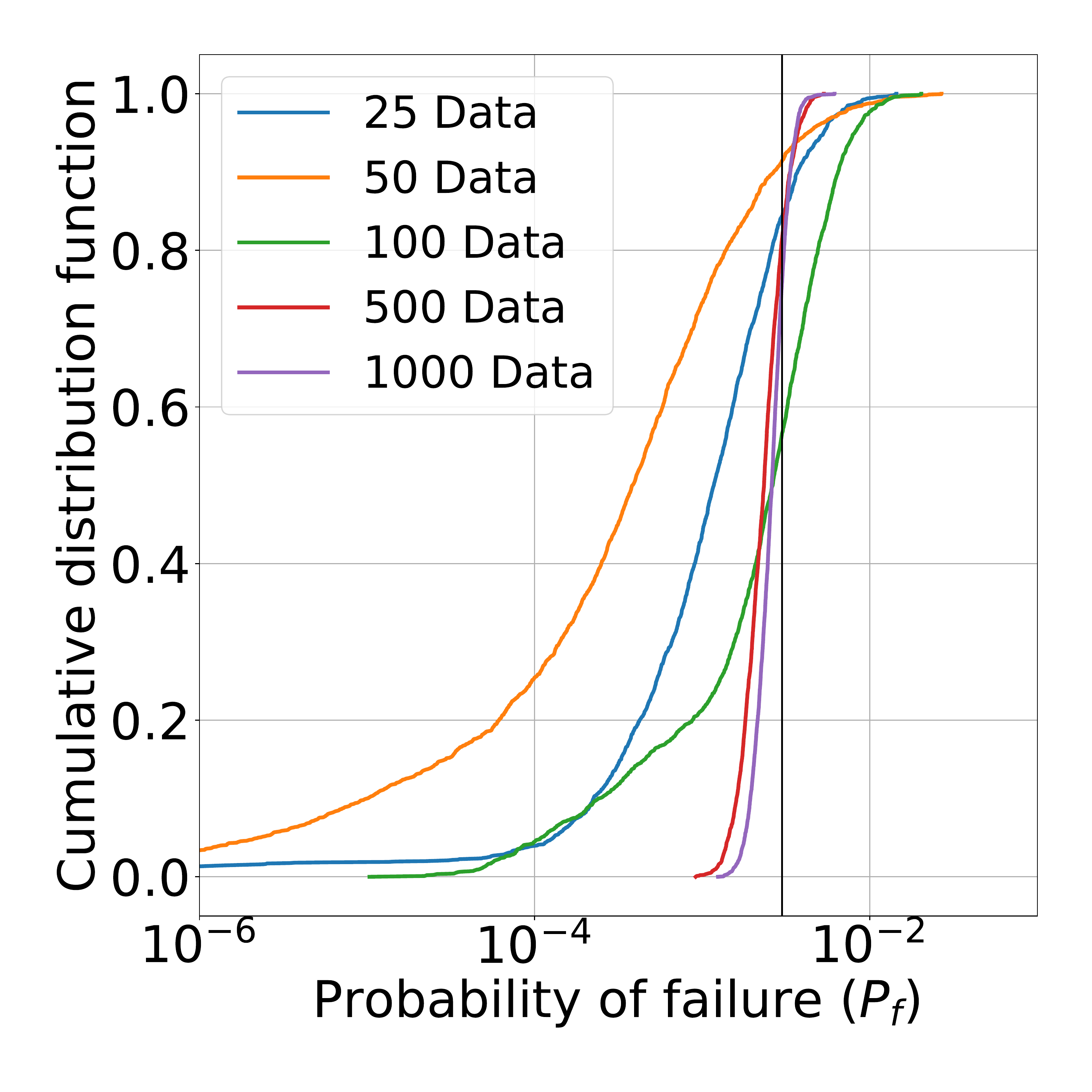}
		\caption{}
	\end{subfigure}
	\caption{Example 1: Comparison of empirical failure probability CDFs from (a)the ``brute force'' approach where SuS is re-run for every candidate distribution, and (b) the re-weighted SuS (iSuS) approach for varying data set size. The true probability of failure is shown by the vertical black line.}
	\label{fig:comp_ex2}
\end{figure}

Next, we run imprecise SuS using both {\it Approach 1} and {\it Approach 2} discussed above. In approach 1, we re-run SuS for each candidate distribution. For the proposed approach, we run SuS once using the optimal sampling density,  calculate the importance weights $w_j$ at each sample point for every distribution model $M_j$ and compute the re-weighted  conditional probabilities $P_{ij}$. Figure  \ref{fig:comp_ex2} shows the resulting empirical cumulative distribution function obtained when we run (a) a single SuS for each candindate distribution model  $M_j$ and (b) the proposed re-weighted  approach, for the different data set sizes.  From this figure we can see that, in both cases uncertainty in the $P_F$ estimates have large uncertainty in $P_F$ estimates for small data sets and convergence towards the true $P_F$ for increasing data set size.  

These results highlight the lack of confidence that we can place in probability of failure estimates when distributions are estimated from very small data sets. Moreover, the proposed approach provides reasonable estimates of this uncertainty with only a single subset simulation.

\subsection{ Example 2: Two degree-of-freedom structure}

\begin{figure}[!ht]
	\centering
	\includegraphics[width=1.0\textwidth]{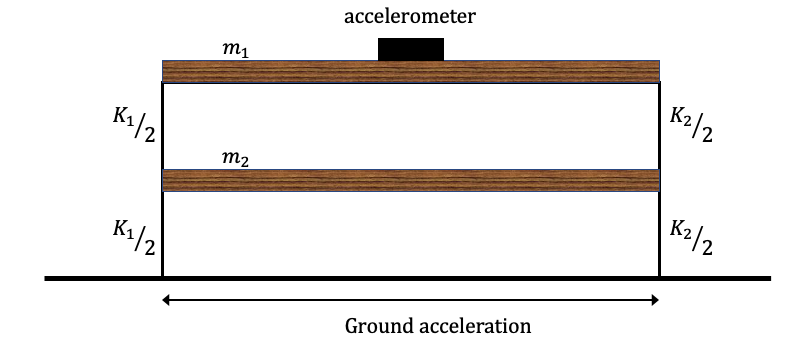}
	\caption{Two-story linear frame structure subjected to ground acceleration.}
	\label{fig:example3}
\end{figure}

\begin{figure}[!ht]
	\centering
	\includegraphics[width=0.5\textwidth]{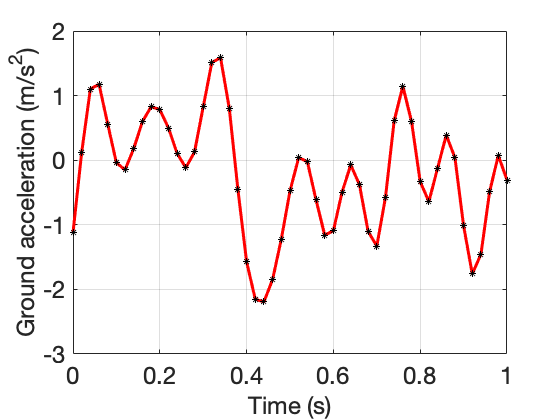}
	\caption{Sample realization of the ground acceleration generated with the spectral representation method. We measure the acceleration for a  period of time $T=1$ sec every $dt=0.2$ sec (a total of 50 time-steps).}
	\label{fig:acceleration}
\end{figure}

\noindent
In the second  example, we study the reliability of  a two-story linear frame structure (see Figure \ref{fig:example3})  subjected to ground acceleration. The problem was originally introduced in \cite{TMCMC}.  In this problem, a low-pass white noise stochastic ground acceleration having a constant power spectral density function $S_f(\omega)=S_0=0.0141$ on the frequency range $\omega \in [0, 35.5]$ rad/sec, is simulated for a short time duration ($T=1$ sec) with the Spectral Representation Method so that only the first mode is excited. 
For simplicity in illustration, we generate a single ground acceleration shown in Figure \ref{fig:acceleration} and perform reliability analysis for the structure subjected to this ground motion. However, the ground motion can be modeled as a stochastic process within this framework by introducing the additional random variables of the stochastic expansion with either known or uncertain distributions.  

We consider three uncertain random variables: the stiffness parameters $K_1$, $K_2$ and the damping ratio
$\xi_1=\xi_2=\xi$ of the two modes. All three random variables are assumed to follow a true lognormal distribution with parameters given in Table \ref{tab:2DOF_params}. The parameters of the correspond Gaussian distributions for these lognormal variables are also provided.
\begin{table}[!ht]
	\centering
	\begin{tabular}{cccccl}\toprule
	  & \multicolumn{2}{c}{\textbf{Lognormal}} & \multicolumn{2}{c}{\textbf{Gaussian}} & \\
	    \toprule
		\textbf{Variable} & \textbf{Mean} & \textbf{std} &  \textbf{Mean} & \textbf{std} & \textbf{Description}   \\ \midrule
		$K_1$   & 1000 & 200 &  0.198 & 980.58 & Stiffness parameter\\
		$K_2$   & 1000 & 200 &  0.198 & 980.58 & Stiffness parameter  \\
		$\xi$   & 0.03 & 0.0045 & 0.149 & 0.029 & Damping ratio \\
		\bottomrule
	\end{tabular}
	\caption{Statistical properties of the  stiffness parameters $K_1$, $K_2$  and the damping ratio $\xi$ for the two-story linear structure.}
\end{table}\label{tab:2DOF_params}
The masses are assumed to be known ($m_1=m_2=1$).   We define failure such that the maximum roof displacement $u_{\max}$ exceeds the threshold 0.022, i.e.,  $P_F = P(u_{\max}\geq 0.022)$.  

We calculate the true probability of failure to be $P_F=2.4 \times 10^{-4}$ using Monte Carlo simulation with $5\times10^4$ samples. Performing SuS with the true distribution parameters, using 1000 samples per subset and conditional probabilities $P_i=0.1$, we find the  probability of failure as $P_F=2.2 \times 10 ^{-4}$, with a coefficient of variation (c.o.v) equal to 0.51.

\subsection*{Imprecise Subset Simulation}
\begin{figure}[!ht]
	\centering
	\begin{subfigure}[t]{0.32\textwidth}
		\centering
		\includegraphics[width=\textwidth]{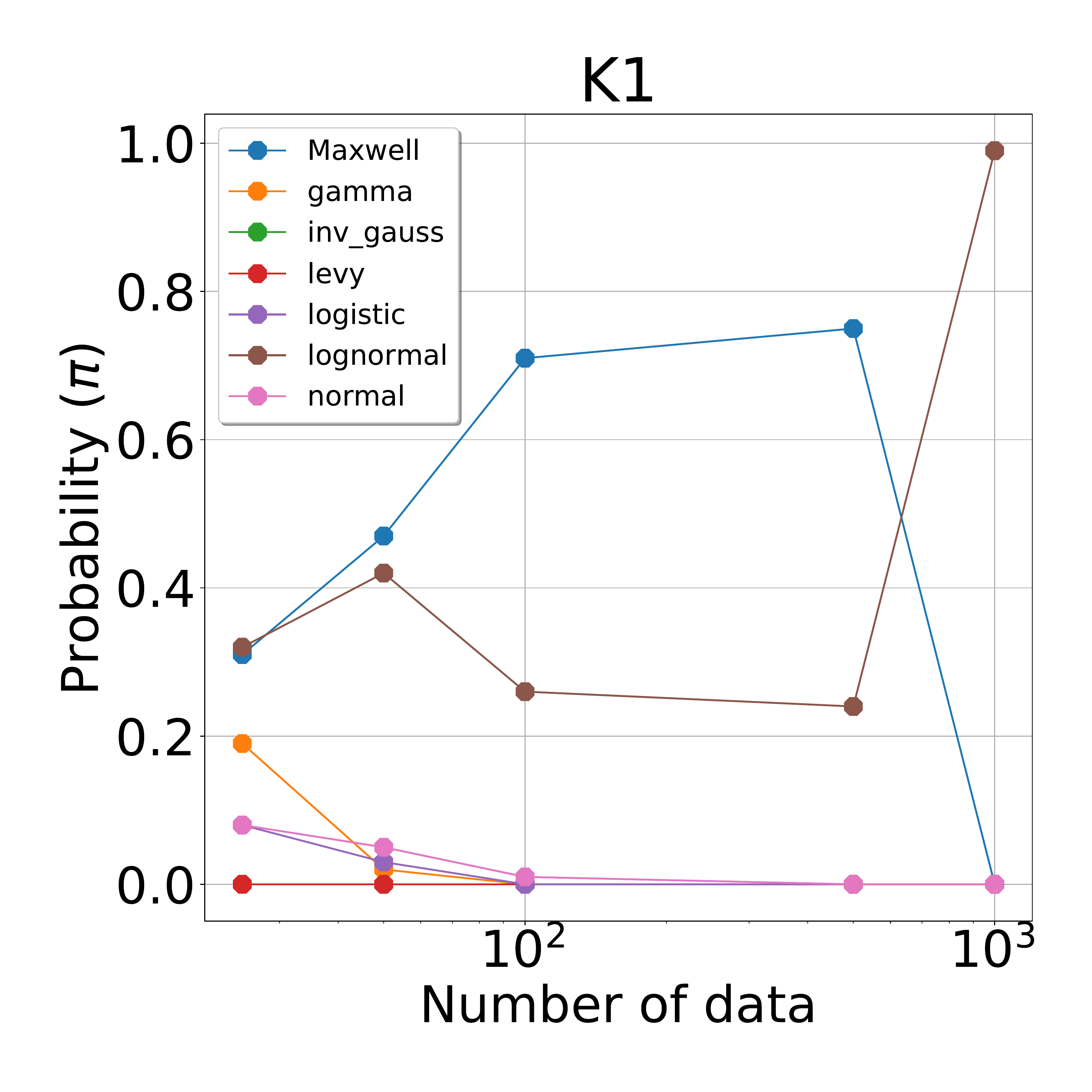}
		\caption{}
	\end{subfigure}
	\begin{subfigure}[t]{0.32\textwidth}
		\centering
		\includegraphics[width=\textwidth]{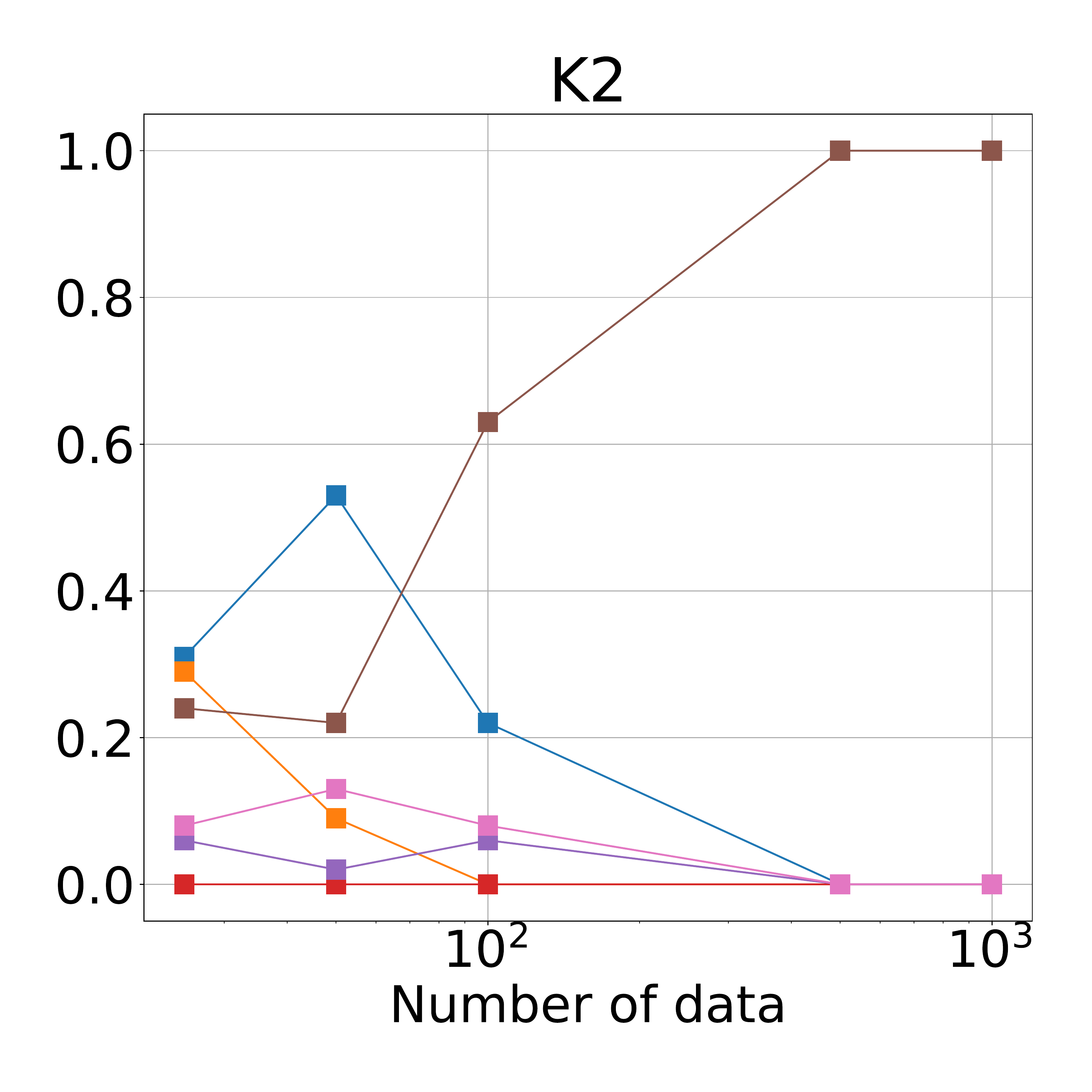}
		\caption{}
	\end{subfigure}
	\begin{subfigure}[t]{0.32\textwidth}
	\centering
	\includegraphics[width=\textwidth]{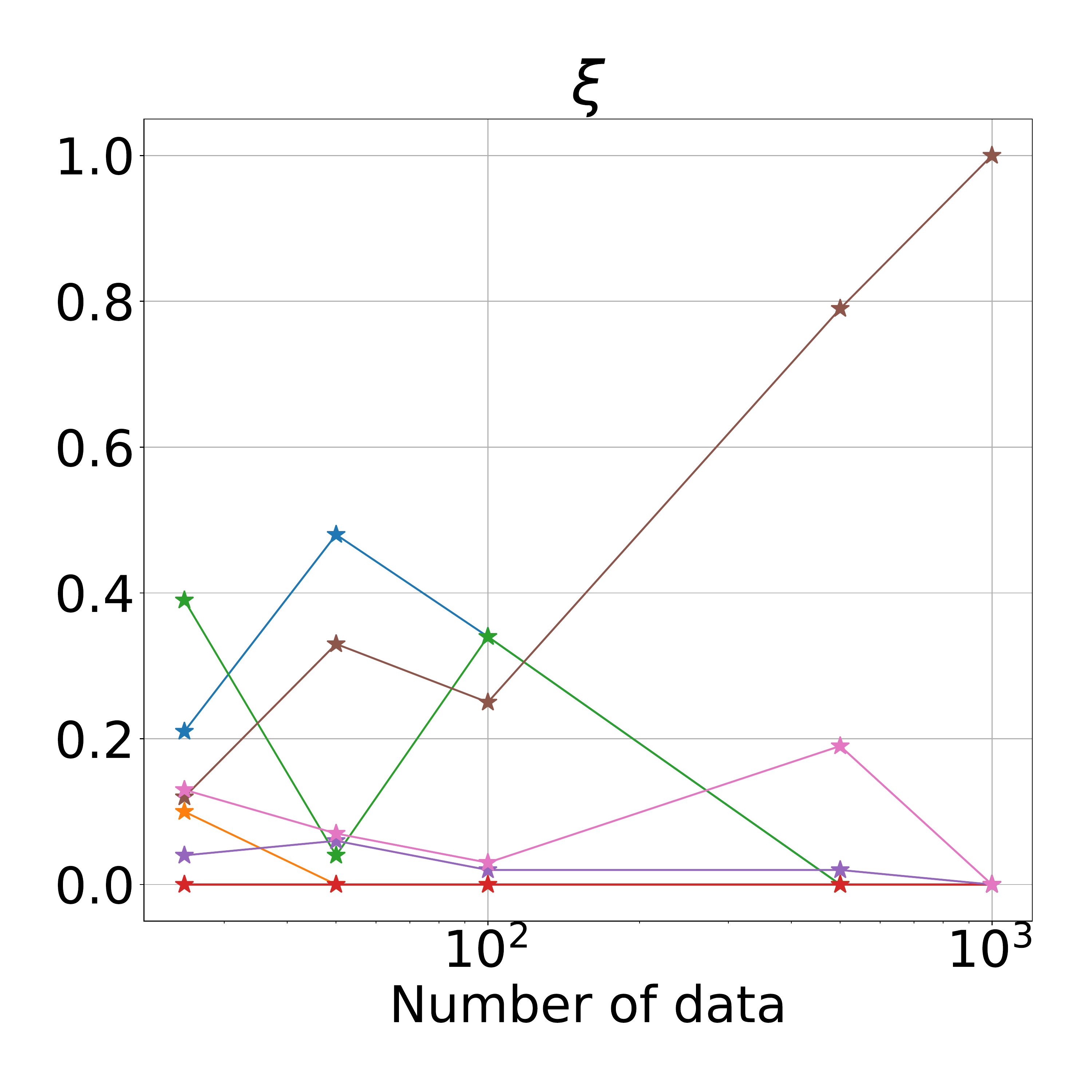}
	\caption{}
\end{subfigure}
	\caption{Example 2: Model probabilities for increasing data set size.}
	\label{fig:Prob_ex1}
\end{figure}

\noindent
Given limited data from which to infer the distributions of $K_1$, $K_2$  and  $\xi$, the multi-model selection framework is applied to select the candidate probability models. Again, we consider the seven families of distributions listed above. Figure  \ref{fig:Prob_ex1} shows the model probabilities of model form as a function of data set size. From this figure, we see that large data sets ($>500$ samples) are necessary to identify the true distribution with high confidence.

\begin{figure}[!ht]
	\centering
	\begin{subfigure}[t]{0.32\textwidth}
		\centering
		\includegraphics[width=\textwidth]{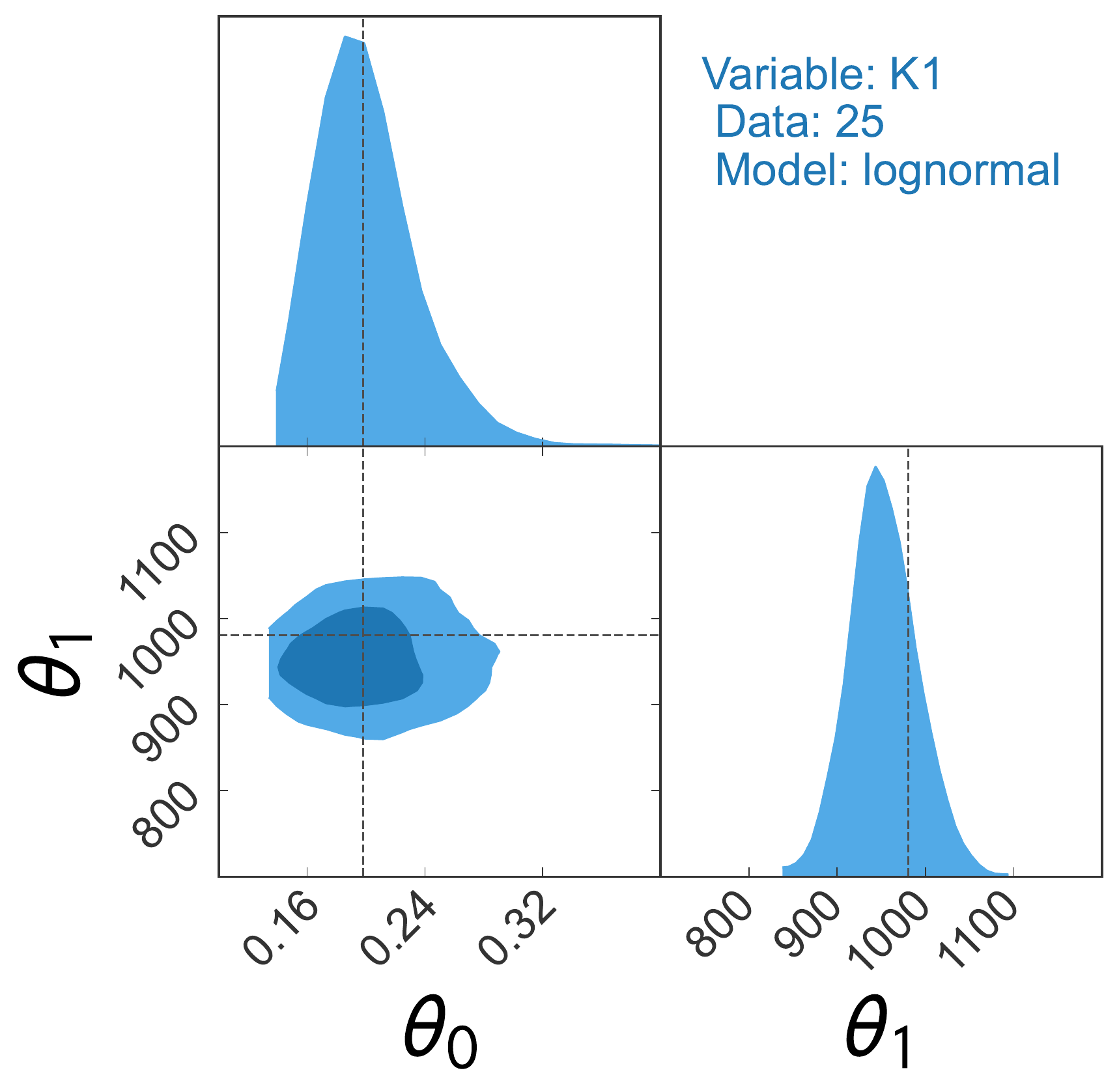}
	\end{subfigure}
	\begin{subfigure}[t]{0.32\textwidth}
		\centering
		\includegraphics[width=\textwidth]{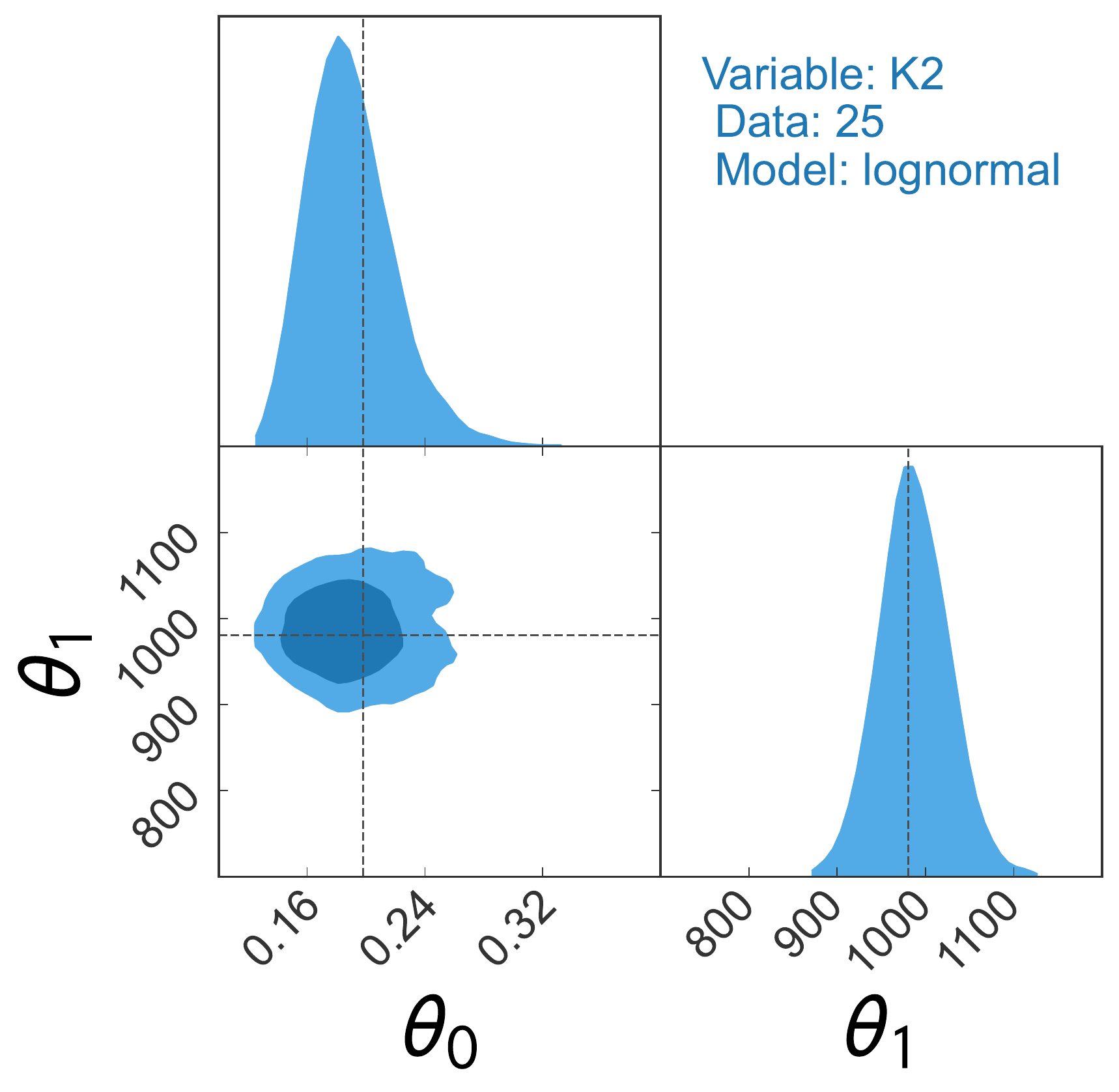}
	\end{subfigure}
	\begin{subfigure}[t]{0.32\textwidth}
	\centering
	\includegraphics[width=\textwidth]{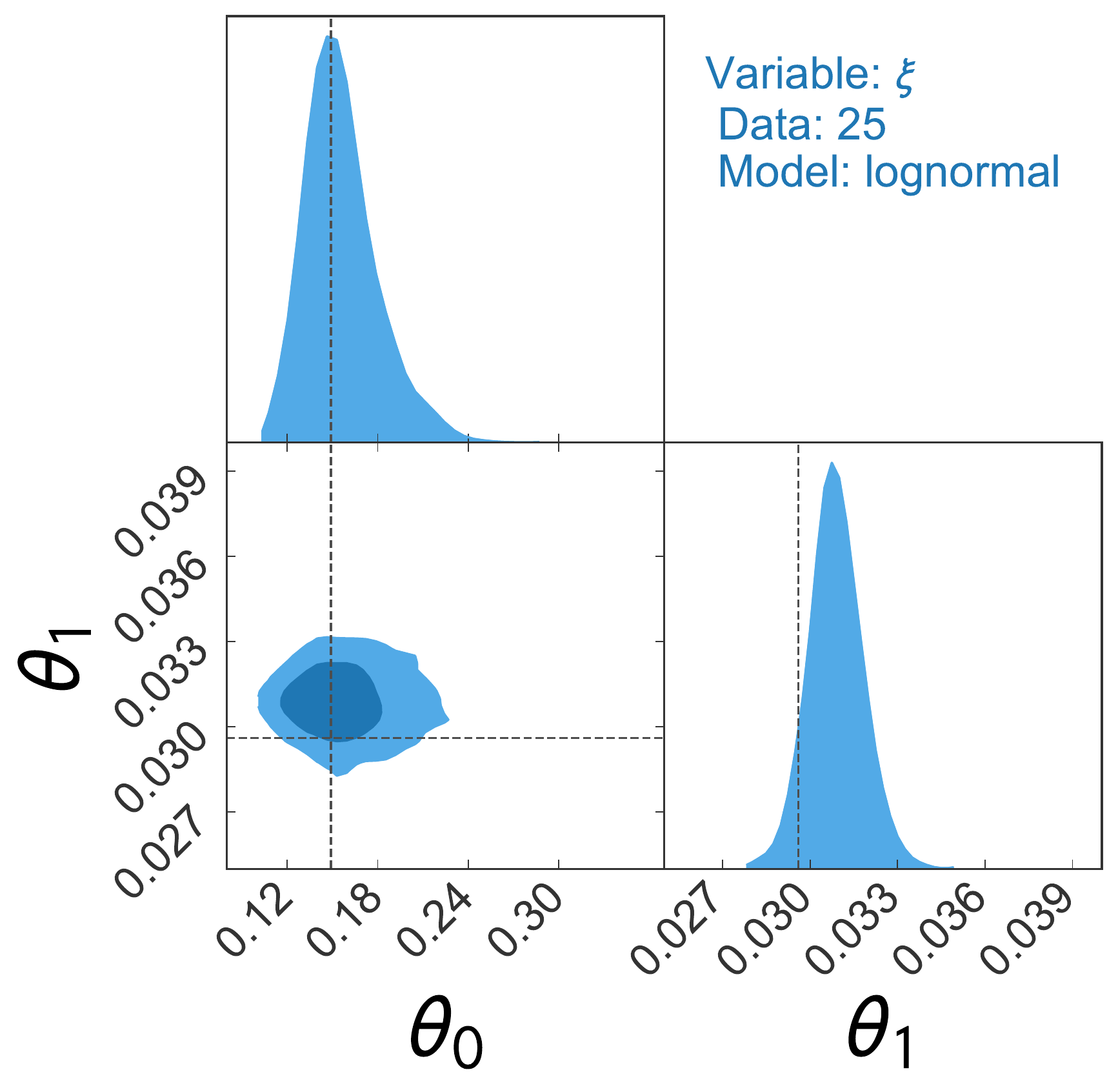}
\end{subfigure}
	\begin{subfigure}[t]{0.32\textwidth}
		\centering
		\includegraphics[width=\textwidth]{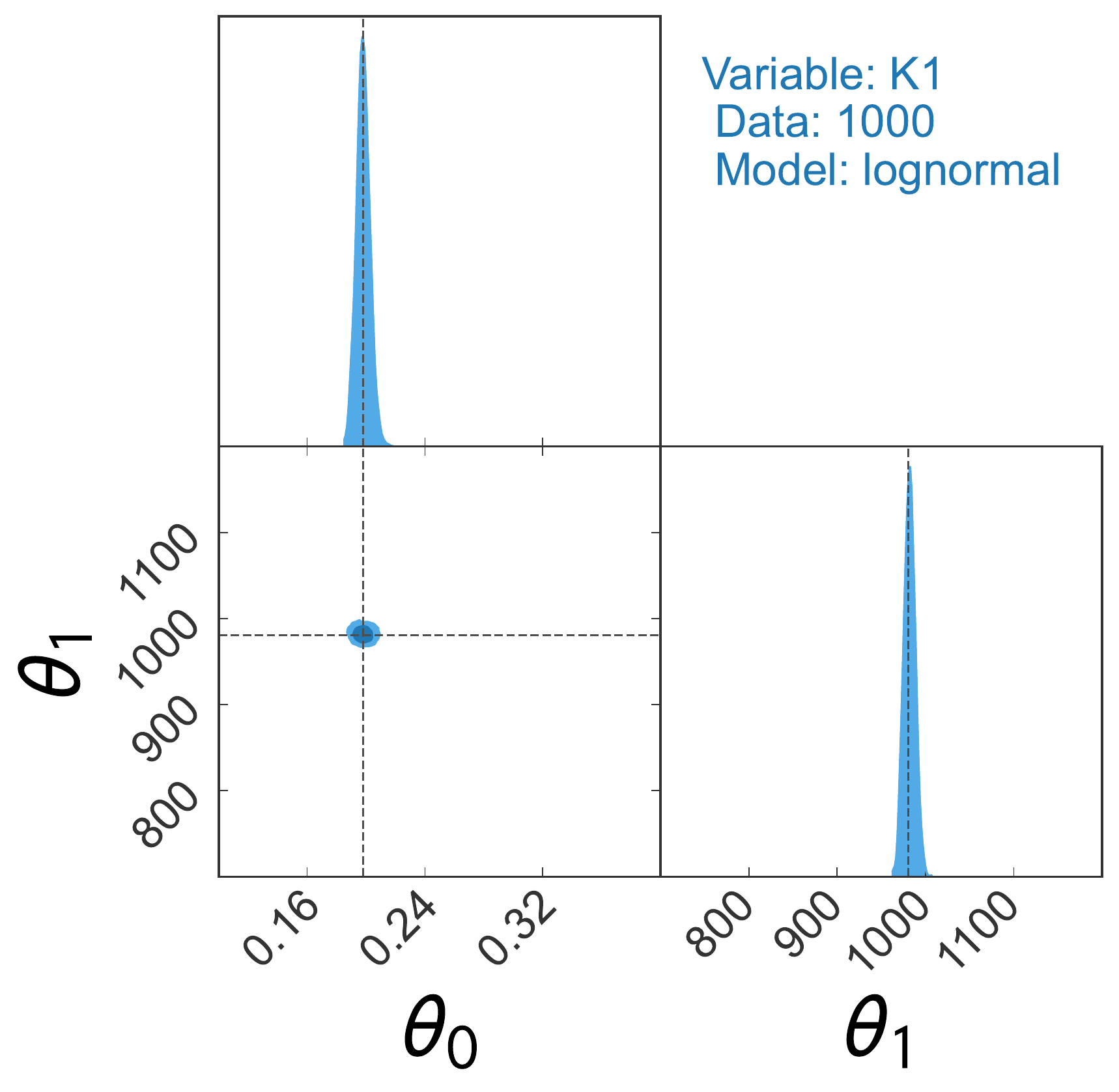}
	\end{subfigure}
	\begin{subfigure}[t]{0.32\textwidth}
		\centering
		\includegraphics[width=\textwidth]{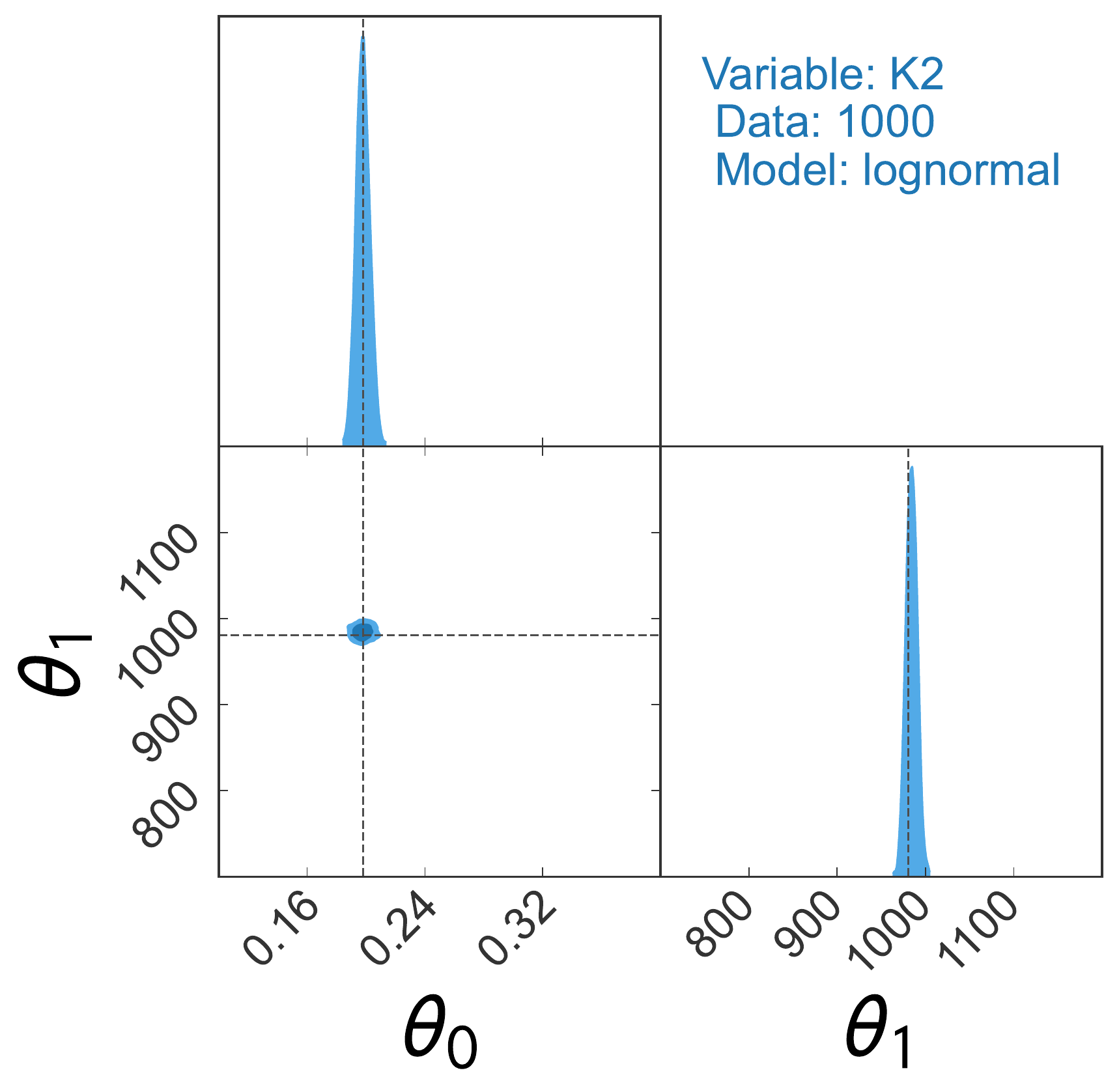}
	\end{subfigure}
	\begin{subfigure}[t]{0.32\textwidth}
		\centering
		\includegraphics[width=\textwidth]{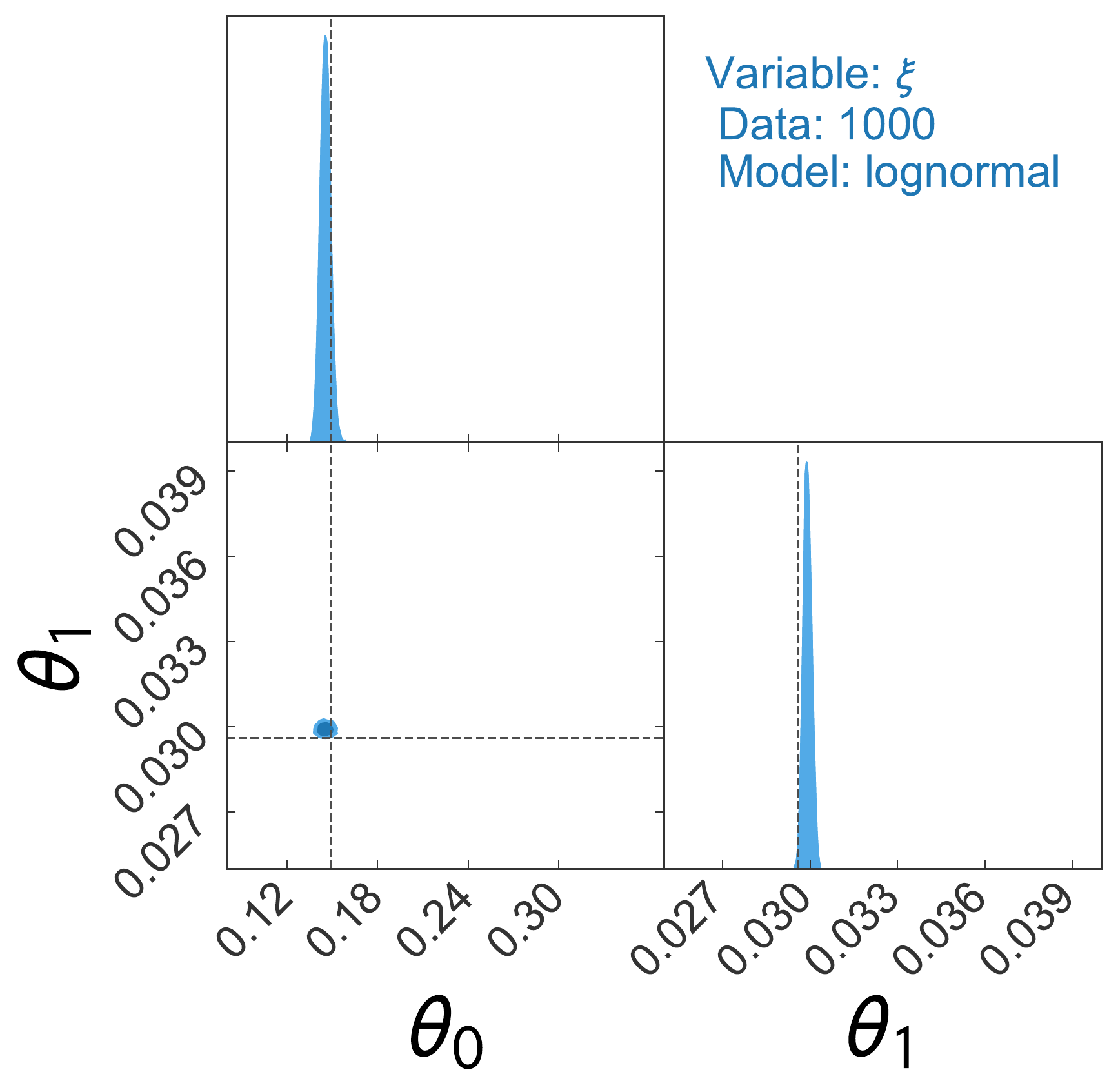}
	\end{subfigure}
\caption{Example 2:  Joint distribution of the underlying Gaussian parameters for the lognormal model for $K_1$ (left), $K_2$ (middle) and  $\xi$ (right) for 25 (top) and 1000 (bottom) data.}
\label{fig:Dists_params_ex3}
\end{figure}

Next, the joint parameter distribution for each of the selected models is determined through Bayesian inference. An example of this inference is shown in Figure \ref{fig:Dists_params_ex3} where, for brevity, only  the joint distributions of the underlying Gaussian model parameters for the lognormal model with 25 and 1000 data are shown. That is,  parameters $\theta_1$ and $\theta_2$ in Figure \ref{fig:Dists_params_ex3} correspond to the mean and standard deviation of the underlying Gaussian variables (see Table \ref{tab:2DOF_params}).  

\begin{figure}[!ht]
	\centering
	\begin{subfigure}[t]{0.40\textwidth}
		\centering
		\includegraphics[width=\textwidth]{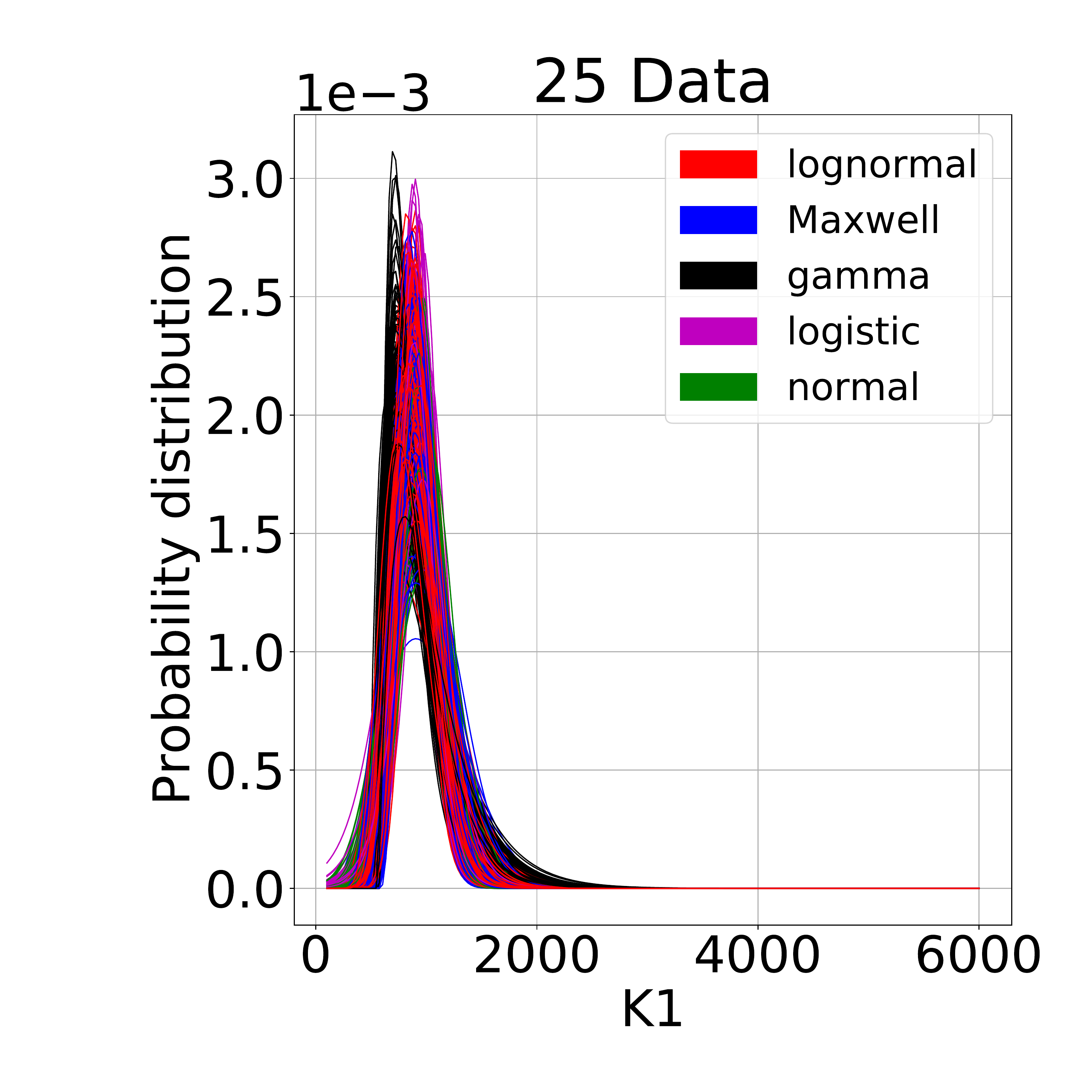}
	\end{subfigure}
	\begin{subfigure}[t]{0.40\textwidth}
		\centering
		\includegraphics[width=\textwidth]{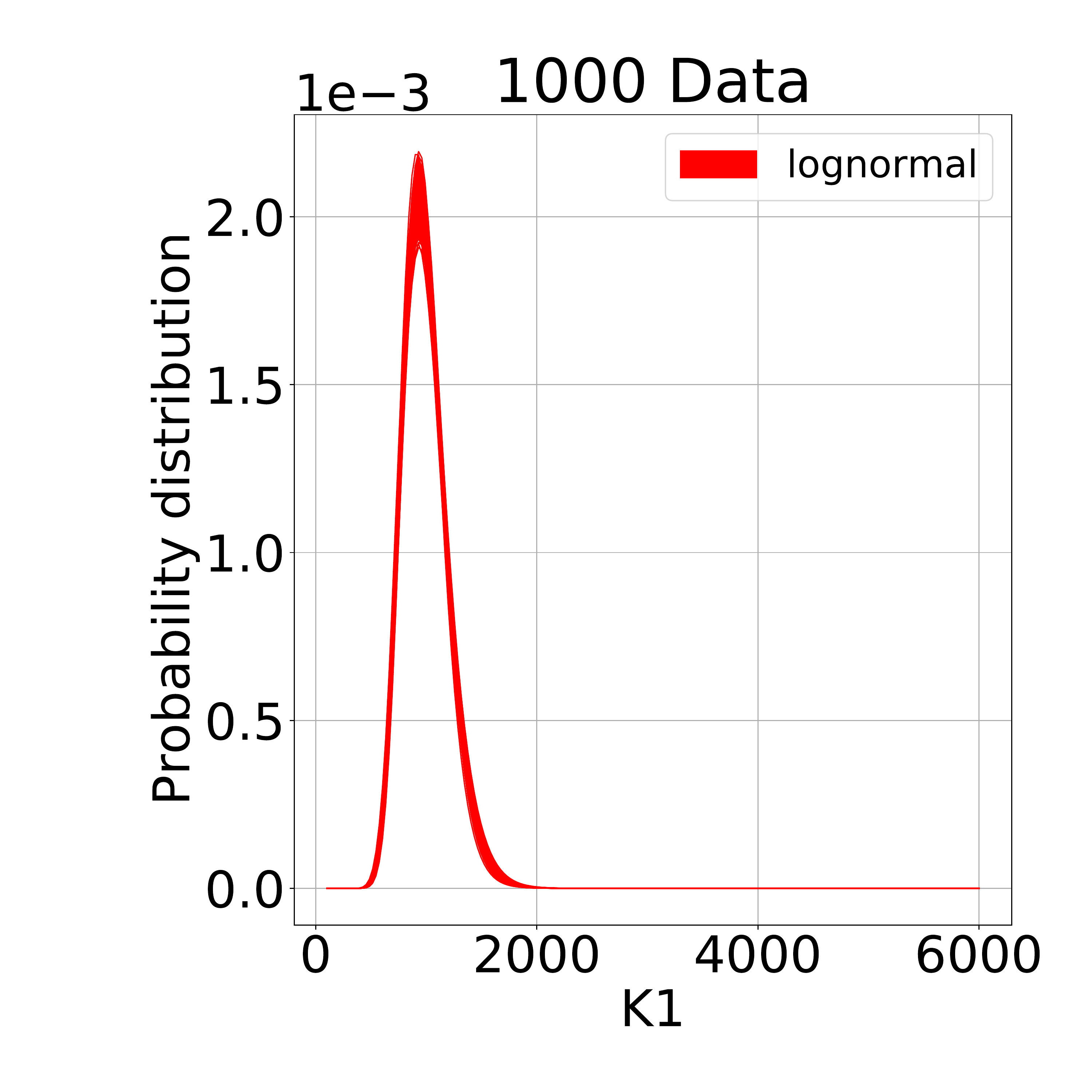}
	\end{subfigure}
	
	\begin{subfigure}[t]{0.40\textwidth}
		\centering
		\includegraphics[width=\textwidth]{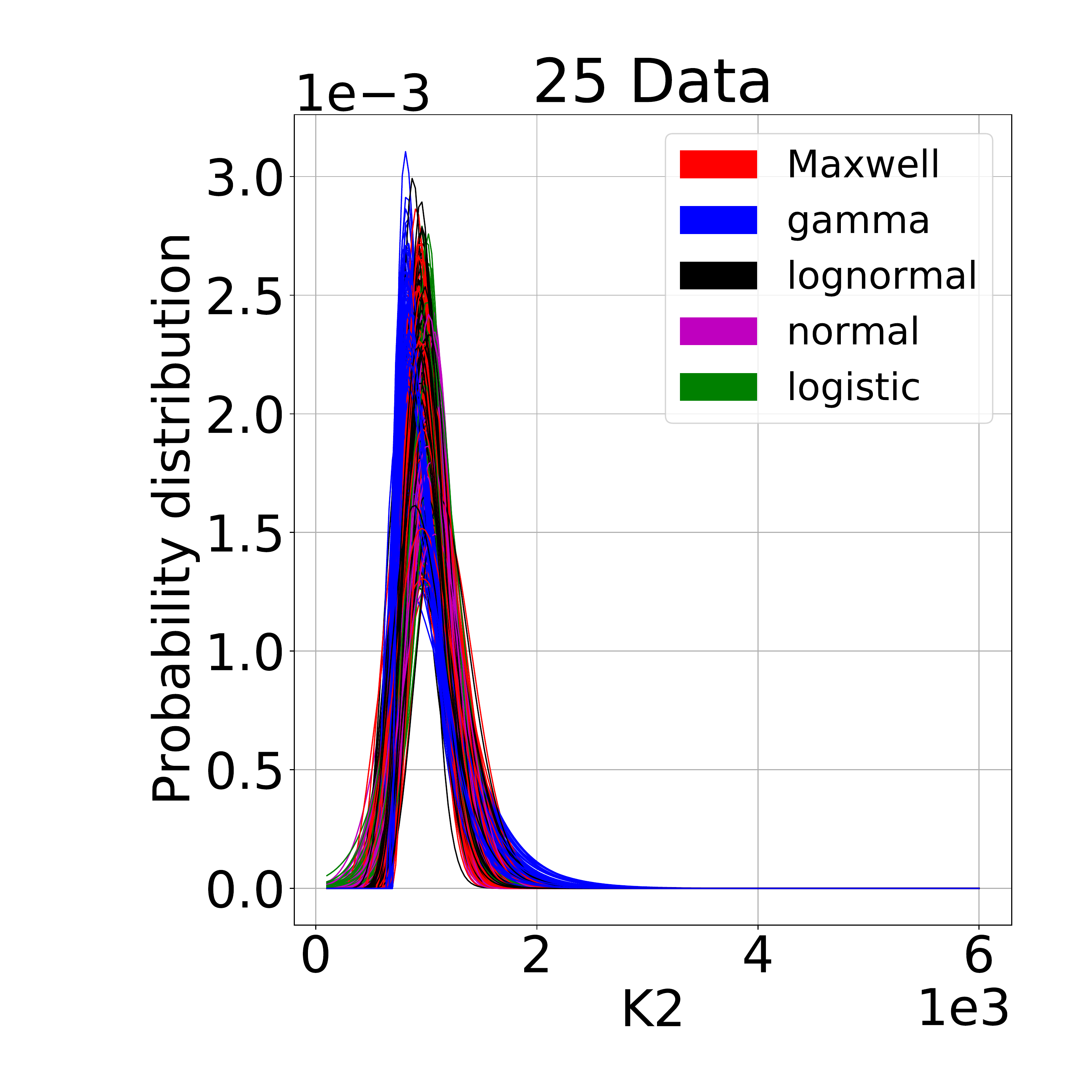}
	\end{subfigure}
	\begin{subfigure}[t]{0.40\textwidth}
		\centering
		\includegraphics[width=\textwidth]{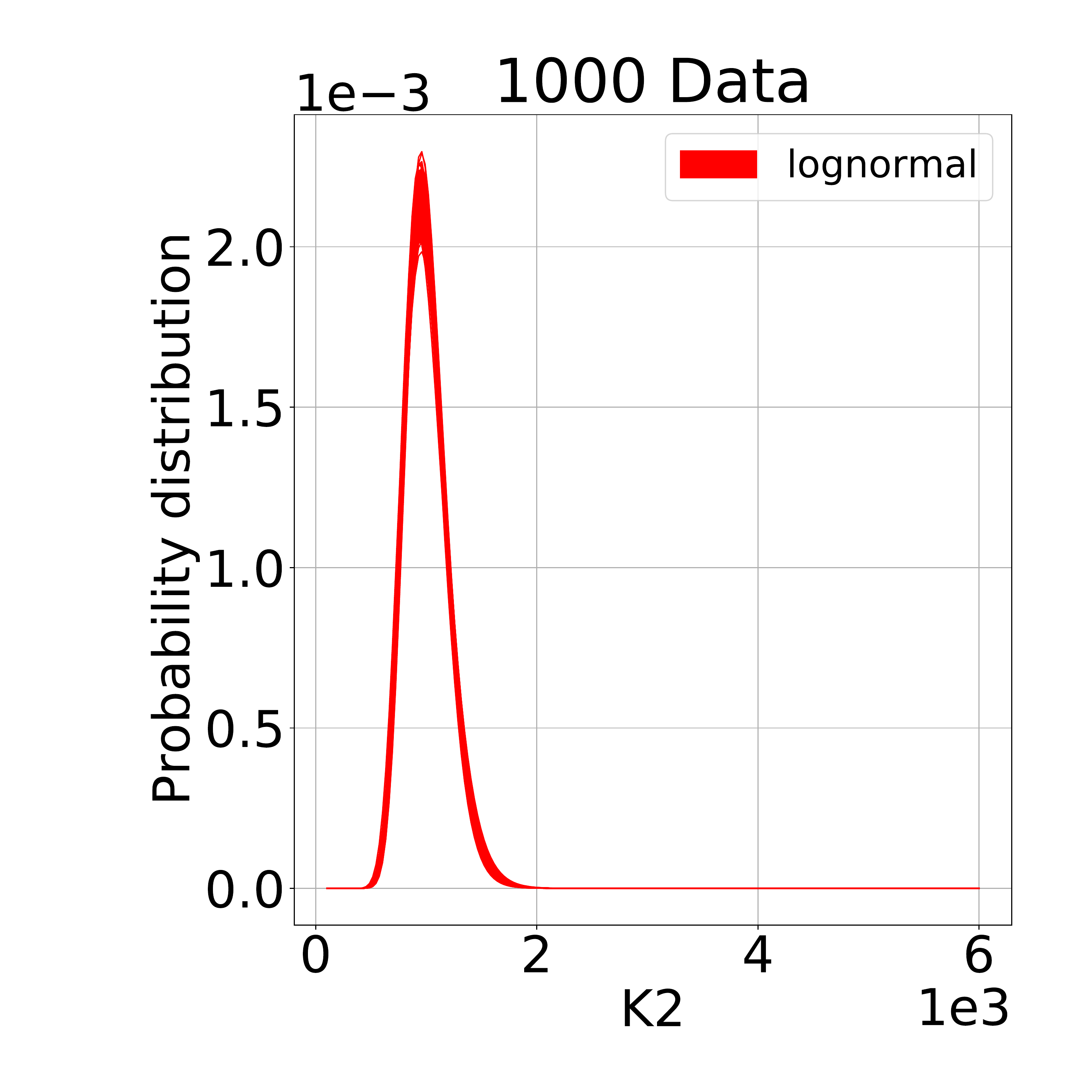}
	\end{subfigure}
	\begin{subfigure}[t]{0.40\textwidth}
	\centering
	\includegraphics[width=\textwidth]{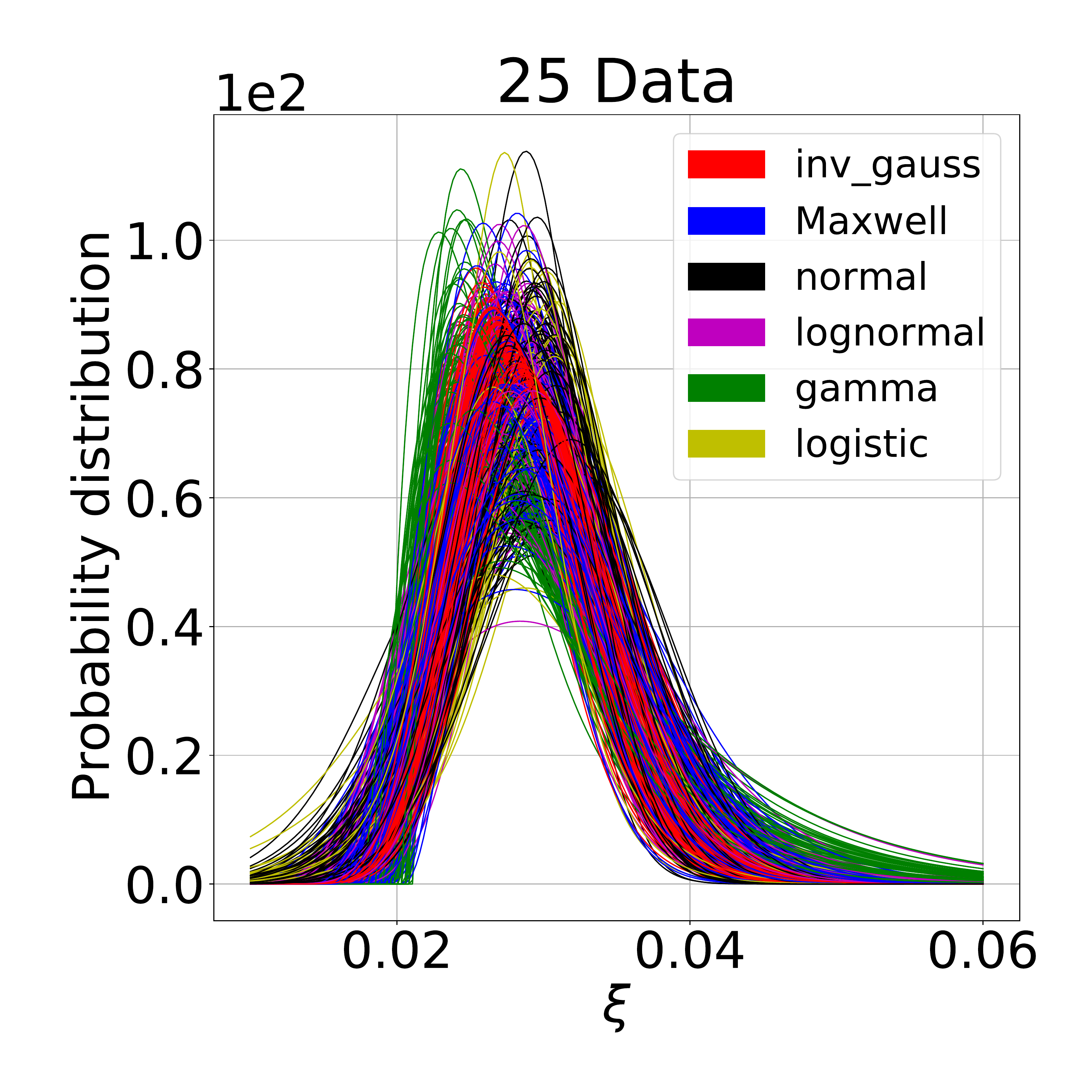}
\end{subfigure}
\begin{subfigure}[t]{0.40\textwidth}
	\centering
	\includegraphics[width=\textwidth]{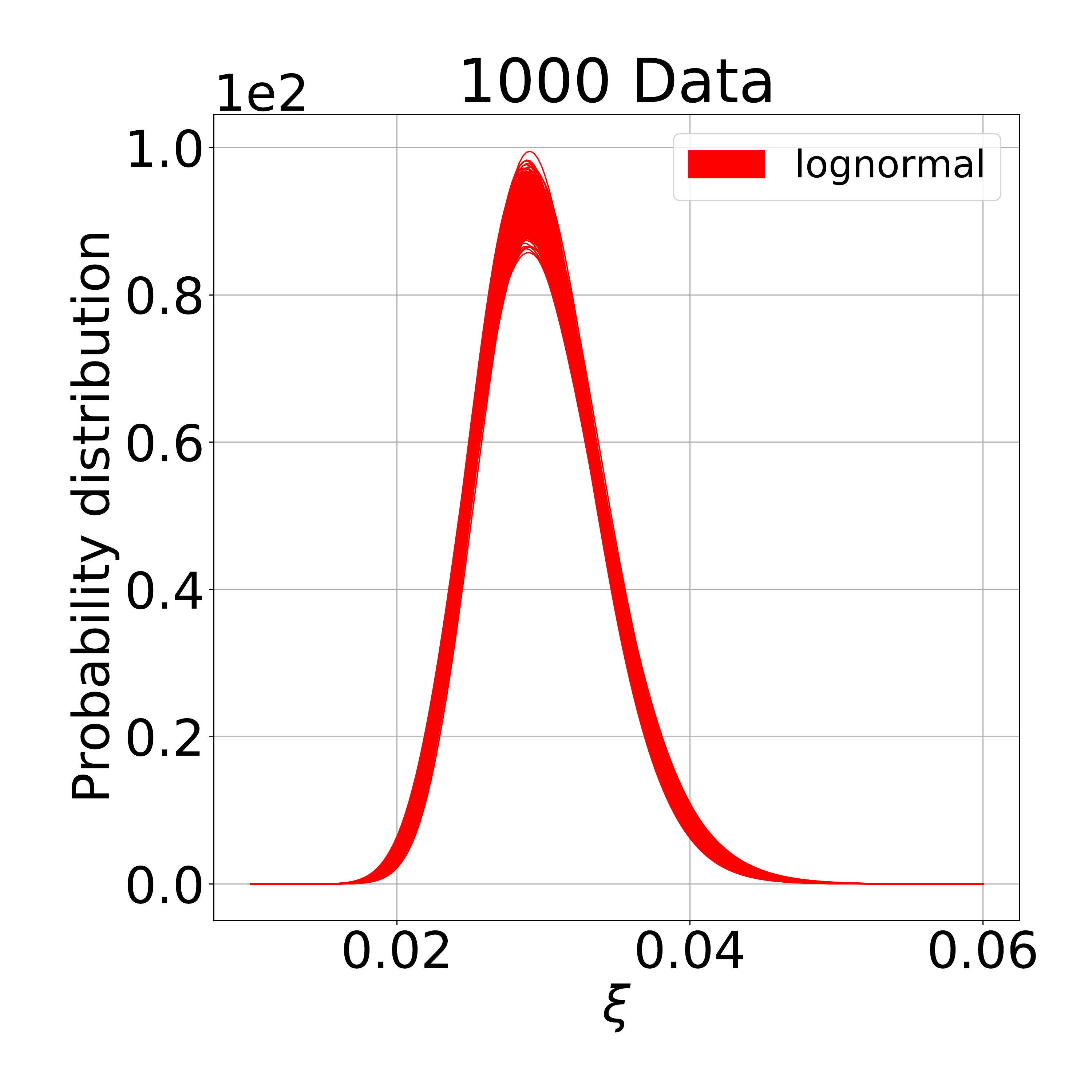}
\end{subfigure}
	\caption{Example 2: Sample sets of probability distributions given 25 and 1000 data for $K_1$ (top), $K_2$ (middle)  and  $\xi$ (bottom).}
	\label{fig:Dists_ex1}
\end{figure}

Figure \ref{fig:Dists_ex1} shows a sample set of candidate distributions for $k_1$, $k_2$  and  $\xi$ for cases where 25 and 1000 data are available for each random variable. Notice that the number of candidate model families reduces and the clouds of candidate distributions that represent the total uncertainty become narrower for increasing data set size.

To assess the influence of model-form and parameter uncertainty on $P_F$ estimates, we apply the proposed iSuS method. First, we calculate the optimal joint density function $q_n^\star(\vec{x})$ from the optimal marginal densities $q_j^\star(\vec{x})$ following the procedure described in  Section 5.  Next, we run SuS once using this distribution  $q_n^\star(\vec{x})$ to identify the optimal conditional performance levels $F_i^{opt}$ and the conditional samples $ \vec{x}_k^{opt}$ at each subest $i$. The importance weights $w_j(\vec{x}_k^{opt})$ are then calculated at each sample point for every distribution model $\mathcal{M}_j$.  Each conditional probability $P_{ij}$ is re-weighted and the probability of failure $P_{Fj}$ for each model in the set is calculated.

\begin{figure}[!ht]
	\centering
	\includegraphics[width=0.8\textwidth]{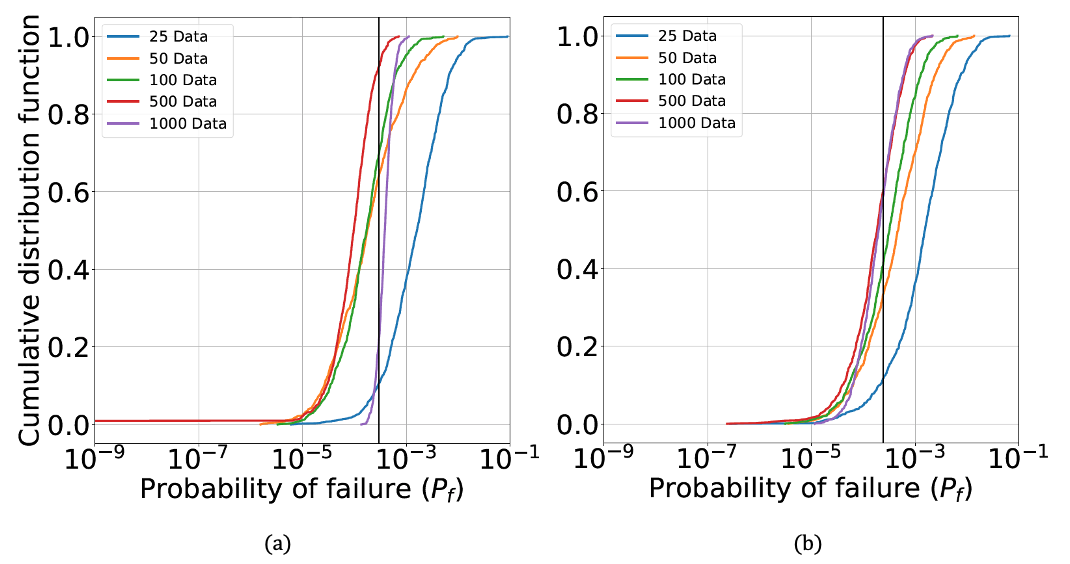}
	\caption{Example 2:  Comparison of empirical failure probability CDFs from (a) the “brute force” approach where SuS is re-run for every candidate distribution, and (b) the re-weighted SuS (iSuS) approach for varying data set size. The true probability failure is shown by a vertical black line.}
	\label{fig:iSuS_cdf}
\end{figure}

Figure \ref{fig:iSuS_cdf} shows the resulting empirical cumulative distribution functions for increasing data set sizes for the case when: (a)  subset simulation is repeated for every candidate model, and  (b) the proposed re-weighting approach. Here, we can see that the proposed iSuS method with sample re-weighting approximates the uncertainty in $P_F$ estimates quite accurately compared to repeated subset simulations, but is dramatically more efficient computationally. The proposed method requires only a single subset simulation plus the nominal cost of re-weighting, whereas the repeated subset simulation approach requires $n_c=1000$ distinct subset simulations that come at huge expense.

\section{Conclusions}

\noindent
This work aims to assess confidence in structural reliability estimates using Subset Simulation (SuS) when data for quantifying input uncertainty are scarce. We emphasize that SuS is very efficient and precise when probability models for input random variables (and hence the conditional levels) are precisely defined.  However, when probability models are uncertain, this induces uncertainty in each conditional level, which further induces uncertainty in the probability of failure estimate and makes the application of SuS challenging.

We propose a method to quantify the confidence in SuS estimates. The proposed scheme leverages a Bayesian-Information Theoretic multimodel inference process for estimating distribution form and parameter uncertainty. This multimodel inference is paired with SuS by first identifying a multimodel set of candidate distributions, solving for an optimal sampling density that best represents this multimodel set, and performing subset simulation using this optimal sampling density. Conditional probablities are then re-weighted according to each distribution in the multimodel set using importance sampling, which results in a statistical set of failure probabilities. This is used to estimate an empirical distribution for the failure probability that can be leveraged to assess confidence in failure probability estimates. The approach is computationally very efficient, requiring only a single subset simulation plus a small cost associated with sample re-weighting, while producing useful probabilistic bounds on reliability estimates.

We demonstrate the proposed method using two examples. The first considers a simple plate buckling problem with non-Gaussian uncertainties inferred from limited material property data. The second problem considers a two-story linear frame structure subjected to ground acceleration with uncertainties in the stiffness and damping that are estimated from small data sets. Both problems illustrate that, when data sets are small, failure probabilities are highly uncertain, sometimes varying by several orders of magnitude.  

\bibliographystyle{elsarticle-num-names}
\bibliography{ImpreciseSuS.bib}

\end{document}